\documentclass[journal]{IEEEtran}
\IEEEoverridecommandlockouts
\usepackage{cite}
\usepackage{amsmath,amssymb,amsfonts}
\usepackage{graphicx}
\usepackage{textcomp}
\usepackage{xcolor}
\def\BibTeX{{\rm B\kern-.05em{\sc i\kern-.025em b}\kern-.08em
    T\kern-.1667em\lower.7ex\hbox{E}\kern-.125emX}}

\usepackage{hyperref}
\usepackage{fancyhdr}    
\usepackage{microtype}  
\usepackage{comment}  

\usepackage{algorithm}
\usepackage{algpseudocode}
\usepackage{lettrine}

\algdef{SE}[SUBALG]{Indent}{EndIndent}{}{\algorithmicend\ }%
\algtext*{Indent}
\algtext*{EndIndent}

\newcommand{\thh}[1]{ {#1^{\rm{th} } } }

\newcommand{\tlambda}[1]{ \mathcal{T}_{#1} }

\newcommand{\yy}{ \mathbf{y} }
\newcommand{\ggb}{ \mathbf{g} }

\newcommand{\bb}{ \mathbf{b} }

\newcommand{\xx}{ \mathbf{x} }
\newcommand{\cc}{ \mathbf{c} }
\newcommand{\vv}{ \mathbf{v} }

\newcommand{\bbig}{ \mathbf{B} }

\newcommand{\cbig}{ \mathbf{C} }
\newcommand{\ybig}{ \mathbf{Y} }
\newcommand{\gbig}{ \mathbf{G} }
\newcommand{\xbig}{ \mathbf{X} }
\newcommand{\vbig}{ \mathbf{V} }

\newcommand{\boldPhi}{ \mathbf{\Phi} }

\newcommand{\varphitilde}{ \varphi^{\text{S}} }
\newcommand{\bbhat}{ \widehat{\bb} }
\newcommand{\bbighat}{ \widehat{\bbig} }

\newcommand{\Nphi}{ C_{\varphi} }
\newcommand{\bpq}{ b_{p,q} }
\newcommand{\complexset}[2]{ \mathbb{C}^{#1 \times #2}  }

\newcommand{\deltaf}{ \Delta f }
\newcommand{\Imatrix}{{ \boldsymbol{\mathrm{I}} }}

\newcommand{\normsmall}[1]{\big\lVert#1\big\rVert}

\newcommand{\vecc}[1]{ {\rm{vec}}\left(#1\right)  }
\newcommand{\veccs}[1]{ {\rm{vec}}\bigg(#1\bigg)  }

\newcommand{\diag}[1]{ {\rm{diag}}\left(#1\right)  }

\newcommand{\lambdamax}{\lambda_{\rm{max}}}
\newcommand{\bbhatls}{\bbhat_{\rm{LS}}}

\newcommand{\bbighatls}{\bbighat_{\rm{LS}}}
\newcommand{\bbighatista}{\bbighat_{\rm{ISTA}}}

\ifCLASSOPTIONcompsoc
\usepackage[caption=false,font=footnotesize]{subfig}
\else
\usepackage[caption=false,font=footnotesize]{subfig}
\fi

\allowdisplaybreaks

\usepackage{scalerel}
\usepackage{tikz}
\usetikzlibrary{svg.path}

\definecolor{orcidlogocol}{HTML}{A6CE39}
\tikzset{
  orcidlogo/.pic={
    \fill[orcidlogocol] svg{M256,128c0,70.7-57.3,128-128,128C57.3,256,0,198.7,0,128C0,57.3,57.3,0,128,0C198.7,0,256,57.3,256,128z};
    \fill[white] svg{M86.3,186.2H70.9V79.1h15.4v48.4V186.2z}
                 svg{M108.9,79.1h41.6c39.6,0,57,28.3,57,53.6c0,27.5-21.5,53.6-56.8,53.6h-41.8V79.1z M124.3,172.4h24.5c34.9,0,42.9-26.5,42.9-39.7c0-21.5-13.7-39.7-43.7-39.7h-23.7V172.4z}
                 svg{M88.7,56.8c0,5.5-4.5,10.1-10.1,10.1c-5.6,0-10.1-4.6-10.1-10.1c0-5.6,4.5-10.1,10.1-10.1C84.2,46.7,88.7,51.3,88.7,56.8z};
  }
}

\newcommand\orcidicon[1]{\href{https://orcid.org/#1}{\mbox{\scalerel*{
\begin{tikzpicture}[yscale=-1,transform shape]
\pic{orcidlogo};
\end{tikzpicture}
}{|}}}}

\usepackage{hyperref} 

\begin{document}

\title{Millimeter-wave Mobile Sensing and Environment Mapping: Models, Algorithms and Validation}

\author{\IEEEauthorblockN{
Carlos Baquero Barneto \orcidicon{0000-0002-0156-935X}\,, \IEEEmembership{Student Member, IEEE},
Elizaveta Rastorgueva-Foi \orcidicon{0000-0002-2576-7078}\,, 
Musa Furkan Keskin \orcidicon{0000-0002-7718-8377}\,, \\
Taneli Riihonen \orcidicon{0000-0001-5416-5263}\,, \IEEEmembership{Member, IEEE}, 
Matias Turunen, 
Jukka Talvitie \orcidicon{0000-0001-7685-7666}\,, \IEEEmembership{Member, IEEE}, \\
Henk Wymeersch \orcidicon{0000-0002-1298-6159}\,, \IEEEmembership{Senior Member, IEEE}, and 
Mikko Valkama \orcidicon{0000-0003-0361-0800}\,, \IEEEmembership{Fellow, IEEE}}

\thanks{This work was partially supported by the Academy of Finland (grants \#315858, \#328214, \#319994, \#323244, \#346622), Nokia Bell Labs, the Doctoral School of Tampere University, Vinnova grant 2018-01929, MSCA-IF grant 888913 (OTFS-RADCOM), the European Commission through the H2020 project Hexa-X (Grant Agreement no. 101015956), and the Finnish Funding Agency for Innovation under the ``RF Convergence'' project.}
\thanks{C.~Baquero Barneto, E.~Rastorgueva-Foi, T.~Riihonen, M.~Turunen, J.~Talvitie and M.~Valkama are with the Unit of Electrical Engineering, Tampere University, 33100 Tampere, Finland (e-mail: mikko.valkama@tuni.fi).}
\thanks{M.~F.~Keskin and H.~Wymeersch are with the Department of Electrical Engineering, Chalmers University of Technology, SE-412 96 Gothenburg, Sweden.}%
\thanks{Measurement data available at {\color{blue}\url{https://doi.org/10.5281/zenodo.4475160}} {[59]}.}%
}

\maketitle

\begin{abstract}
Integrating efficient connectivity, positioning and sensing functionalities into 5G New Radio (NR) and beyond mobile cellular systems is one timely research paradigm, especially at mm-wave and sub-THz bands. In this article, we address the radio-based sensing and environment mapping prospects with specific emphasis on the user equipment (UE) side. We first describe an efficient $\ell_1$-regularized least-squares (LS) approach to obtain sparse range--angle charts at individual measurement or sensing locations. For the subsequent environment mapping, we then introduce a novel state model for mapping diffuse and specular scattering, which allows efficient tracking of individual scatterers over time using interacting multiple model (IMM) extended Kalman filter and smoother. {Also the related measurement selection and data association problems are addressed.} We provide extensive numerical indoor mapping results at the 28~GHz band deploying OFDM-based 5G NR uplink waveform with 400~MHz channel bandwidth, covering both accurate ray-tracing based as well as actual RF measurement results. The results illustrate the superiority of the dynamic tracking-based solutions, compared to static reference methods, while overall demonstrate the excellent prospects of radio-based mobile environment sensing and mapping in future mm-wave networks.
\end{abstract}

\begin{IEEEkeywords}
5G New Radio (NR), 6G, joint communications and sensing, mobile radar, RF convergence, indoor mapping, mm-waves, sub-THz, OFDM radar. 
\end{IEEEkeywords}


\section{Introduction}
\label{sec:Introduction}

\lettrine[lines=2]{\textbf{F}} \enskip IFTH generation (5G) New Radio (NR) mobile cellular systems provide large improvements in terms of, e.g, peak data rates, network capacity, number of connected devices, and radio access latency, compared to earlier Long-Term Evolution (LTE) based networks \cite{2019:B:Holma:5GTechnology}. While the primary purpose of mobile networks is efficient data connectivity, they can also facilitate terrestrial positioning service -- with ambitious accuracy requirements of around one meter in 5G NR \cite{journal_47,journal_49,conference_2,3GPPTS22261}. Additionally, to further leverage the capabilities of millimeter-wave (mm-wave) transceiver systems, the radio-frequency (RF) convergence paradigm refers to integrating also radio-based sensing capabilities to the networks and the corresponding user devices \cite{journal_1, journal_2, journal_15, journal_21, journal_23}. To this end, the development of joint communication and sensing (JCAS) systems that can perform both functionalities while sharing the same transmit waveforms and hardware platforms is a very timely research area \cite{journal_3, journal_63, myPaper_2_TMTT19,myPaper_6_WCM20}, with applications, e.g., in vehicular systems and indoor mapping \cite{journal_48,9330512,journal_3,journal_5,journal_6}.  
The large channel bandwidths available at mm-wave frequencies, together with highly directional antenna arrays, form the technical basis for high-accuracy radio positioning and sensing \cite{journal_4,conference_2,myPaper_3_Asilomar19}. While 3GPP 5G NR supports currently channel bandwidths up to 400~MHz \cite{3GPPTS38104}, further increased channel bandwidths are expected through the NR evolution towards the sub-THz regime \cite{9275614}, paving eventually the way to future 6G networks \cite{9330512,8766143}. 

\begin{figure*}[t]
    \centering
    \subfloat[]{\includegraphics[width=0.5\textwidth]{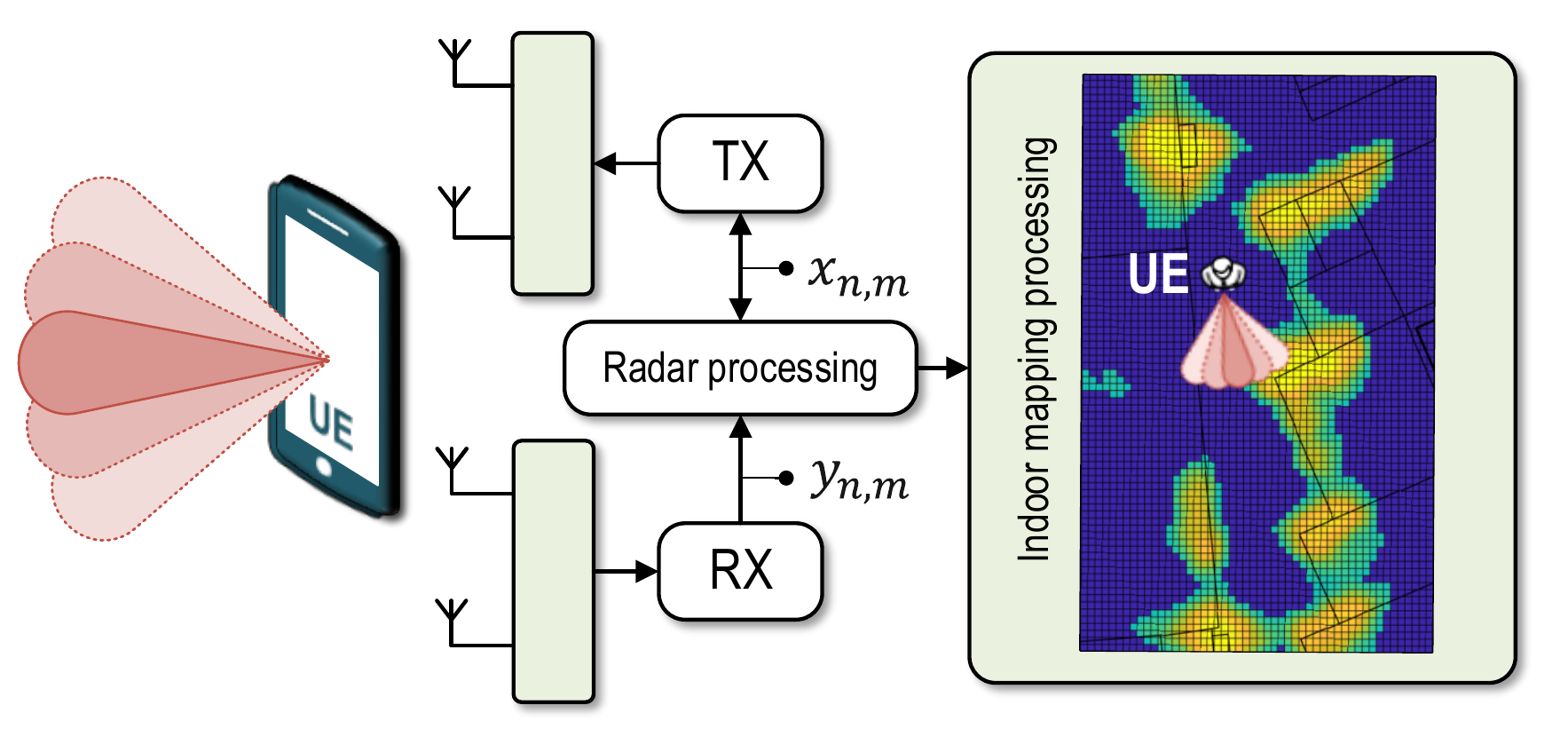}
    \label{fig:blockDiagram}}
    \subfloat[]{\includegraphics[width=0.45\textwidth]{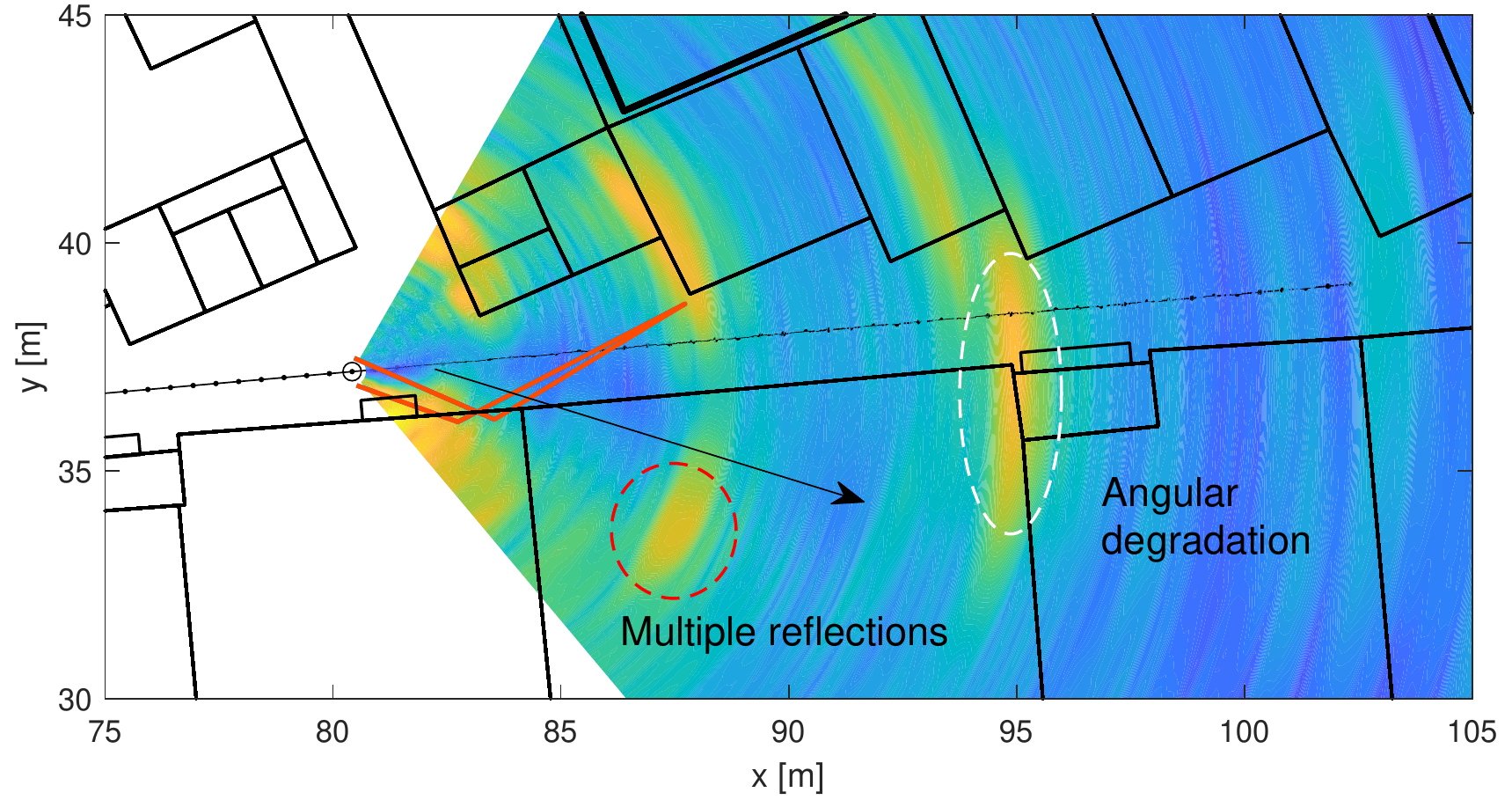}
    \label{fig:challenges}}
    \caption{(a) Considered joint communication and sensing/radar scenario in mm-wave 5G NR network, for user terminal based mapping applications. (b) Example measurement based illustration of selected indoor mapping challenges related to angular degradation as well as to multi-bounce phenomenon at 28~GHz.}
    \label{fig:systemConcept}
\end{figure*}

In general, in typical \emph{outdoor} mm-wave scenarios with densely deployed network nodes or radio heads, the user equipment (UE) is likely to have line-of-sight (LoS) connection to the network while also being within the coverage range of multiple network nodes -- both being issues that further facilitate high-accuracy positioning \cite{journal_47,journal_49,conference_2}. However, in \emph{indoor} deployment scenarios, the propagation characteristics and thus the positioning become much more challenging due to the high path loss as well as the reduced penetration and diffraction \cite{journal_40,journal_52}. 
To address such challenges, the concept of simultaneous localization and mapping (SLAM) has been studied. To this end, in \cite{journal_74, conference_50, conference_51, journal_76, conference_52, conference_53, journal_77, journal_78, journal_50, journal_51,conference_34}, different SLAM algorithms aim to estimate the UE position and the environment scatterer positions by exploiting the multipath component (MPC) information from physical or fixed anchors, e.g., 5G NR base stations.

In \cite{journal_50}, a map-free indoor localization method using ultra-wideband (UWB) mm-wave signals is proposed, utilizing the estimated scatterer positions of the environment as virtual anchors to obtain the UE location. {However, a snap-shot approach is taken, thus ignoring the correlation in sequences of measurements obtained by
moving sensors.}
The work in \cite{journal_51}, in turn, described a SLAM architecture for mm-wave localization through obstacle detection and dimensioning for indoor environments.
Furthermore, a message passing-based estimator which jointly estimates the position and orientation of the UE while sensing the environment's reflectors is presented in \cite{conference_34}. {The works \cite{journal_51} and \cite{conference_34} do not consider the actual MPC estimation and do not take into account the different nature of the specular reflections and diffuse scattering in the actual mapping task.}

In contrast to the traditional SLAM methods which fuse information from fixed anchors and moving agents, alternative techniques consider \textit{user-centric} approaches where localization and mapping functionalities are implemented at each agent independently, {without the need for fixed infrastructure} \cite{journal_75}. In this case, each moving agent is equipped with transmit and receive antennas, similar to a monostatic radar system, that measures the MPCs and performs SLAM accordingly.
In \cite{journal_75}, {it is assumed that a reference map or floorplan is available, hence focusing mostly on UE localization, while also a recursive update method to account for the uncertainties of the floorplan in user-centric SLAM system is proposed. Only ranging or delay measurements are considered in \cite{journal_75} while the environment related state model comprises essentially of wall segments.}
In \cite{journal_4}, a personal mobile radar concept for environment sensing or mapping is studied, building on large antenna arrays at mm-waves. {The mapping related state variables comprise a collection of radar cross-section (RCS) values for a predeﬁned grid of
geographical cells covering uniformly the overall considered area. Such an approach leads easily to thousands of state variables, and hence large computational complexity.}
In \cite{journal_36}, a JCAS system is pursued for smart home scenarios, investigating different integration approaches for traditional sensing and WiFi devices.
In general, due to the notable challenges in performing RF measurements at mm-wave bands, only  very few works have supported their positioning/sensing/mapping solutions with empirical results to assess and demonstrate the performance with real mm-wave transceiver and antenna hardware \cite{journal_50,journal_52,journal_53, conference_37, journal_81}.

In this article, we describe {a novel application framework for} \textit{user-centric} mm-wave indoor mapping systems, propose associated signal processing methods in 5G NR UE context, {assuming a known agent trajectory, and perform validation with ray-tracing and RF measurements}. 
In the considered approach, illustrated conceptually in Fig.~\ref{fig:systemConcept}, the UE senses the surrounding environment through orthogonal frequency-division multiplexing (OFDM)-based beamformed uplink transmissions while then observing and collecting the target reflections {with synchronous receiver beamforming patterns}.  This is followed by range--angle processing, while the corresponding range--angle charts are then further post-processed in the subsequent dynamic mapping stage. 
{The considered radio-based sensing and mapping concept exploits the UE mobility, which collects correlated sensing measurements at different locations to better reconstruct the environment}. We specifically address the moving nature of the sensing and measurement device, in the form of advanced tracking-based dynamic mapping solutions, incorporating both specular reflections and diffuse scattering. {With the proposed tracking approach, besides mapping of fixed structures, it is in general possible to track moving targets, adapt to the changes in the environment, as well as predict the mobility of a tracked target.}

With respect to the overall state-of-the-art literature in the field, the contributions and novelty of the article can be summarized as follows:
\begin{itemize}
    \item An efficient $\ell_1$-regularized least-squares (LS) {formulation} to obtain sparse range--angle charts at individual measurement or sensing locations, expressed for OFDM systems, is {provided}; 
    \item {A novel state model for mapping diffuse and specular scattering is introduced, which allows efficient tracking of individual scatterers over time using interacting multiple model (IMM) extended Kalman filter/smoother;}
    \item Multi-bounce or double-reflection challenge of practical complex environments is addressed {through the available received signal strength (RSS) measurements and applicable pathloss models}, and corresponding novel measurement selection and data association methods are devised;
    \item Comprehensive numerical indoor mapping results at 28~GHz band deploying OFDM-based 5G NR uplink waveform with 400~MHz channel bandwidth are provided, covering both accurate ray-tracing based as well as actual RF measurement results.
\end{itemize}

The rest of this article is organized as follows. The fundamental system model for collecting beamformed sensing measurements and the associated measurement system geometry are described in Section II. The LS-based range--angle processing methods {for individual sensing locations} are addressed and derived in Section III, while Section IV presents the proposed tracking-based dynamic mapping solutions {for a sequence of sensing locations}, respectively. In Section V, the evaluation scenario, ray-tracing environment and the experimental RF measurement environment and equipment are described. Obtained numerical results are provided and analyzed in Section VI, while the conclusions are provided in Section VII. {Finally, further details of the LS-based range--angle processing solution are described in the Appendices A and B.}

\section{System Model}
\label{sec:systemModel}
\subsection{Basics}
The sensing and mapping functionality of the NR UE builds on the OFDM-based uplink transmit waveform whose complex I/Q samples are fully known in the device. 
This is conceptually illustrated in Fig.~\ref{fig:systemConcept}\subref{fig:blockDiagram}. 
Furthermore, the mm-wave UE is assumed to be equipped with {a directive antenna system for both transmit (TX) and receive (RX) functions, which allows for accurate angular measurements to map the environment. In the following system model description, for notational convenience, uniform linear arrays (ULAs) are considered for both the TX and RX sides. However, it is noted for clarity that directive horn antennas are used in the later RF measurements due to the available hardware limitations.}
Additionally, in this work, we specifically focus on sensing the azimuth domain to create a 2D map of the environment. However, the same principles can be applied for sensing in the elevation domain and subsequently for extracting 3D maps.

For given sensing direction, we assume that the UE transmits a sequence of $M$ beamformed OFDM symbols on $N$ active subcarriers, with $x_{n,m}$ denoting the transmitted frequency-domain symbol at $n^{\text{th}}$ subcarrier in $m^{\text{th}}$ OFDM symbol. 
{The presented system model assumes a static channel across the $M$ transmitted symbols which can be extended to consider moving targets as shown in \cite{braun2014ofdm, myPaper_2_TMTT19}.}
To this end, the frequency-domain OFDM-based radar model with $K$ targets or {MPCs} reads \cite{braun2014ofdm,journal_13,myPaper_2_TMTT19}
\begin{equation}
\label{eq:RX_signal}
    y_{n,m} = \mathbf{w}_\text{RX}^H \mathbf{A}_{\text{RX},n}
    \boldsymbol{\Gamma}_{n}
    \mathbf{A}_{\text{TX},n}^H 
    \mathbf{w}_\text{TX}
    x_{n,m} + v_{n,m},
\end{equation}
where $v_{n,m}$, $\mathbf{w}_\text{TX}\!\in\!\mathbb{C}^{N_{\text{TX}} \times 1}$ and $\mathbf{w}_\text{RX}\!\in\!\mathbb{C}^{N_{\text{RX}} \times 1}$ are the complex Gaussian noise and the TX and RX array beamforming weights, respectively. Moreover, $N_{\text{TX}}$ and $N_{\text{RX}}$ denote the number of antenna elements in the TX and RX array, and $\mathbf{A}_{\text{TX},n}\!\in\! \mathbb{C}^{N_{\text{TX}}\times K}$ and $\mathbf{A}_{\text{RX},n}\!\in\! \mathbb{C}^{N_{\text{RX}}\times K}$ are the steering matrices for the TX and RX array, respectively.
The multipath coefficients of a sensed environment with $K$ target reflections are modelled through ${\boldsymbol{\Gamma}_{n} = \text{diag}(\gamma_{0,n}, \ldots, \gamma_{K-1,n})} \!\in\! \mathbb{C}^{K\times K}$ that is a diagonal matrix with its $k^{\text{th}}$ element being of the form 
\begin{align}\label{eq_gammakn}
    {\gamma_{k,n}  = \sqrt{\frac{\lambda_{n}^2 \sigma^{\textrm{RCS}}_k}
        {\left (  4\pi\right )^3 d_k^4}}
        e^{-j2 \pi \tau_k (f_\text{c}+n\Delta f )}}.
\end{align}
In above, $\tau_k$, $d_k$, $\sigma^{\textrm{RCS}}_k$ and {$f_\text{c}$} model the $k^{\text{th}}$ target's propagation delay, propagation distance, RCS and the {system carrier frequency}, respectively, stemming from the well-known \textit{radar range equation} \cite{braun2014ofdm}. furthermore, $\lambda_{n}$ refers to the wavelength of the $n^{\text{th}}$ subcarrier while $\Delta f$ denotes the subcarrier spacing of the OFDM waveform. 

Under the ULA assumption, the TX array steering vector for a $k^{\text{th}}$ target with azimuth angle $\varphi^{\text{TX}}_k$ can be expressed at subcarrier $n$ as
\begin{equation}
    \mathbf{a}_{\text{TX},n}(\varphi^\text{TX}_k) =
    \left [ 
    1, e^{j{\Upsilon}_n(\varphi^\text{TX}_k)}
    ,\ldots ,
    e^{j(N_{\text{TX}}-1){\Upsilon}_n(\varphi^\text{TX}_k)}
    \right ]^{T},
\end{equation}
where ${\Upsilon}_n(\varphi^\text{TX}_k) = 2\pi \frac{\upsilon }{\lambda_n}\sin{(\varphi^\text{TX}_k)}$ is the corresponding electrical angle of departure (AoD) for the $k^{\text{th}}$ target, wherein $\upsilon$ is the antenna element separation. 
The RX steering vector $\mathbf{a}_{\text{RX},n}(\varphi^\text{RX}_k)$ and the corresponding electrical angle of arrival (AoA) ${\Upsilon }_n(\varphi^\text{RX}_k)$ are obtained similarly.
As a result, the set of steering vectors for the considered $K$ target reflections can be expressed with compact matrix notation as $\mathbf{A}_{\text{TX},n}\!=\![\mathbf{a}_{\text{TX},n}(\varphi^\text{RX}_0), \ldots,\mathbf{a}_{\text{TX},n}(\varphi^\text{RX}_{K-1})]$ and $\mathbf{A}_{\text{RX},n}\!=\![\mathbf{a}_{\text{RX},n}(\varphi^\text{RX}_0), \ldots,\mathbf{a}_{\text{RX},n}(\varphi^\text{RX}_{K-1})]$ which are deployed in~(\ref{eq:RX_signal}).

\begin{figure}[t!]
        \centering
        \includegraphics[width=0.7\columnwidth]{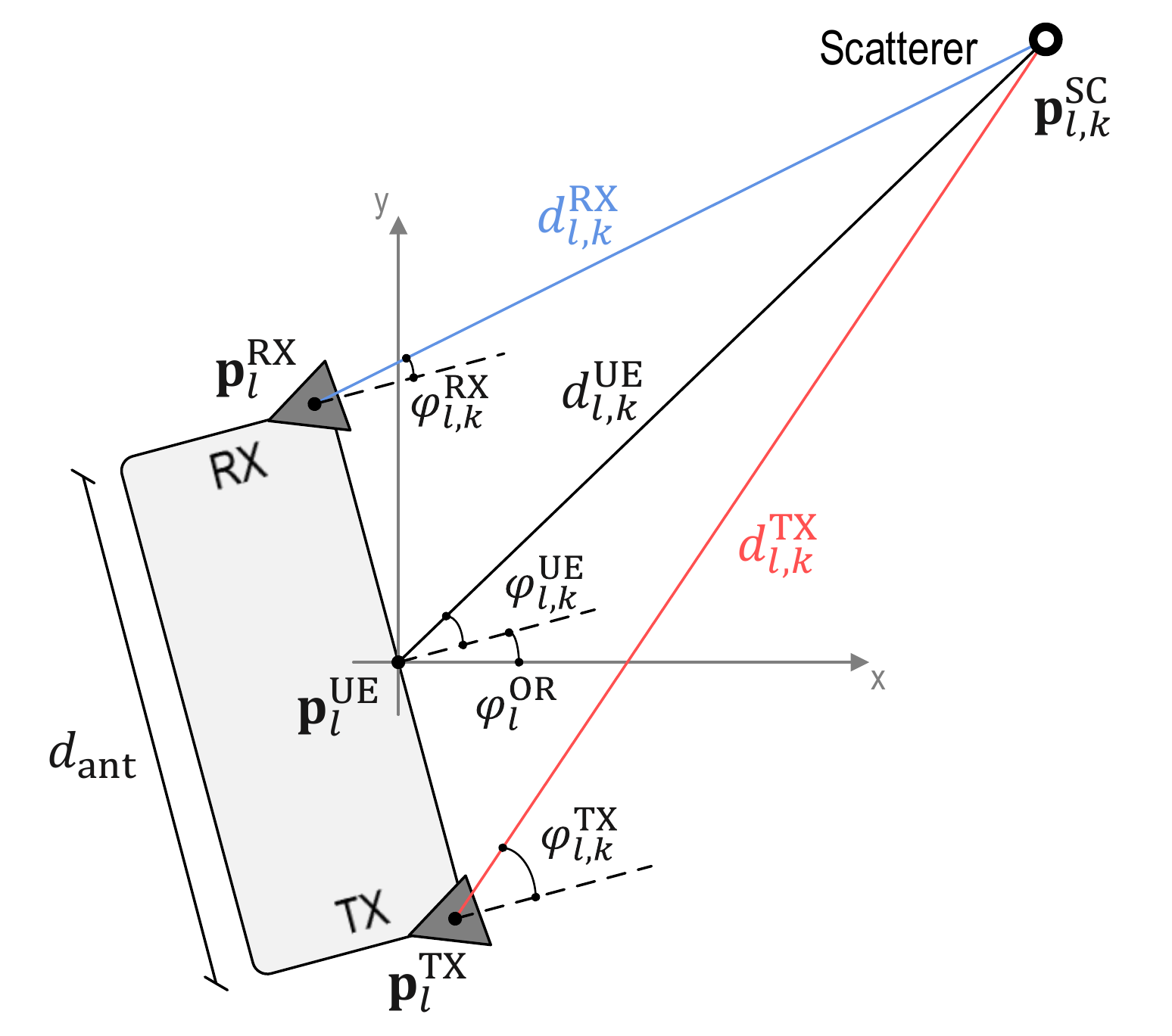}
        \caption{Geometry of the considered sensing setup, where the illumination and measurement system located at ${\mathbf{p}^{\text{UE}}_{l} = (x_{l}^{\text{UE}},y_{l}^{\text{UE}}})$ with orientation $\varphi^{\text{OR}}_l$ senses the environment, containing a target at distance $d^{\text{UE}}_{l,k}$ with respect to the UE location.
        {The TX and RX antenna systems are located at $\mathbf{p}^{\text{TX}}_{l}$ and $\mathbf{p}^{\text{RX}}_{l}$, respectively.}
        }
        \label{fig:ellipseModel}
\end{figure}

\subsection{Geometry}
Next, we consider the specific characteristics related to the fundamental system geometry shown in Fig.~\ref{fig:ellipseModel} 
where the sensing device is located at position ${\mathbf{p}^{\text{UE}}_{l} = (x_{l}^{\text{UE}},y_{l}^{\text{UE}}})$ with $l$ denoting the sensing instance or location index. 
{Later in this work, in Section IV, we consider the fact that the UE is moving, hence leading to a sequence of sensing locations ${\mathbf{p}^{\text{UE}}_{l} = (x_{l}^{\text{UE}},y_{l}^{\text{UE}}})$,  
$l = 1,...,L$. However, here and in the following Section III, an arbitrary given sensing location is considered.} 

Like illustrated in the figure, we allow for separate TX and RX arrays in our system model, to relax the self-interference challenge in OFDM radars \cite{myPaper_2_TMTT19,journal_11}. 
Therefore, with separate TX array position ${\mathbf{p}^{\text{TX}}_{l} = (x_{l}^{\text{TX}},y_{l}^{\text{TX}}})$ and RX array position ${\mathbf{p}^{\text{RX}}_{l} = (x_{l}^{\text{RX}},y_{l}^{\text{RX}}})$, the physical propagation delay of the $k^{\text{th}}$ target reflection $\tau_{l,k} = (d_{l,k}^{\text{TX}} + d_{l,k}^{\text{RX}})/c_0$ is determined cumulatively according to the distance between the TX array and target position $d_{l,k}^{\text{TX}}$, and the distance between the RX array and target position $d_{l,k}^{\text{RX}}$. Consequently, the distance (i.e., range) between the UE position $\mathbf{p}^{\text{UE}}_{l}$ and the $k^{\text{th}}$ target position $\mathbf{p}^{\text{SC}}_{l,k}$ with azimuth angle $\varphi^{\text{UE}}_{l,k}$ can be expressed as
\begin{align}\label{eq_UEdistance}
d_{l,k}^{\text{UE}} = \rho(\tau_{l,k},\varphi^{\text{UE}}_{l,k})=\frac{\sqrt{(c_0\tau_{l,k})^2-d_{\text{ant}}^2}}{2\sqrt{1-\left(\frac{d_{\text{ant}}}{c_0\tau_{l,k}}\cos(\varphi^{\text{UE}}_{l,k})\right)^2}},
\end{align}
where $c_0$ is the speed of light and $d_{\text{ant}}$ corresponds to the separation between TX and RX arrays. For the special case of co-located or joint TX-RX antenna system, $d_{\text{ant}}=0$ while the above expression is written in a general form. 
Furthermore, for a given physical propagation delay $\tau_{l,k}$ and the system geometry shown in Fig.~\ref{fig:ellipseModel}, the transmit angle $\varphi^{\text{TX}}_{l,k}$ -- and similarly the receive angle -- can be described as
\begin{align}\label{eq_TXRXangle}
\begin{split}
\varphi^{\text{TX}}_{l,k} = \operatorname{atan2}(y_{l}^{\text{UE}}-y_{l}^{\text{TX}}+\rho(\tau_{l,k},\varphi^{\text{UE}}_{l,k})\sin(\varphi^{\text{UE}}_{l,k}),\\
x_{l}^{\text{UE}}-x_{l}^{\text{TX}}+\rho(\tau_{l,k},\varphi^{\text{UE}}_{l,k})\cos(\varphi^{\text{UE}}_{l,k})),
\end{split}
\end{align}
where $\rho(\tau_{l,k},\varphi^{\text{UE}}_{l,k})$ is defined in (\ref{eq_UEdistance}) and $\operatorname{atan2}(y,x)$ is a four-quadrant inverse tangent function.

Assuming then that the distances of the considered 
scatterers satisfy the far-field condition,
we can express the TX and RX gain patterns associated to a target or scatterer $k$ as
\begin{align}
    g_{\text{TX},n} (\varphi^{\text{TX}}_{l,k}) & =   \mathbf{a}_{\text{TX},n} ^H (\varphi^{\text{TX}}_{l,k})\mathbf{w}_\text{TX} , \nonumber \\ 
    g_{\text{RX},n} (\varphi^{\text{RX}}_{l,k}) & = 
    \mathbf{w}_\text{RX} ^H
    \mathbf{a}_{\text{RX},n}(\varphi^{\text{RX}}_{l,k})    .
\end{align}
For simplicity, in the rest of the paper, we assume frequency-flat gain patterns with 
$g_{\text{TX},n} (\varphi^{\text{TX}}_{l,k}) = g_{\text{TX}} (\varphi^{\text{TX}}_{l,k})$ and $g_{\text{RX},n} (\varphi^{\text{RX}}_{l,k}) = g_{\text{RX}} (\varphi^{\text{RX}}_{l,k})$.
Finally, the combined TX-RX antenna pattern is defined as
\begin{align}\label{eq_pattern_g}
    g(\varphi^{\text{UE}}_{l,k}) =  g_{\text{RX}}(\varphi^{\text{RX}}_{l,k})   g_{\text{TX}}(\varphi^{\text{TX}}_{l,k}) =
    \mathbf{w}_\text{RX} ^H \mathbf{a}_{\text{RX}}(\varphi^{\text{RX}}_{l,k}) 
    \mathbf{a}_{\text{TX}} ^H (\varphi^{\text{TX}}_{l,k})\mathbf{w}_\text{TX}.
    \end{align}

For generality, we also note that in the JCAS literature, different approaches have been considered to optimize the TX and RX beamforming weights, $\mathbf{w}_\text{TX}$ and $\mathbf{w}_\text{RX}$, see, e.g., \cite{journal_7,journal_27,journal_35,journal_37}. Specifically,
different advanced techniques {enabling the transmission of} multiple simultaneous beams have been proposed in these recent works.
In this paper, however, we consider more ordinary single-beam approach due to its implementation simplicity, while defer the potential multi-beam considerations to our follow-up future work.

\section{Range--Angle Processing}
\label{sec:rangeAngle_processing}
In this section, we start by extending the basic signal model in \eqref{eq:RX_signal} to the case of multiple sensing directions {for range--angle charting}. Then, we introduce a novel formulation of the OFDM radar range--angle processing problem for the proposed environment mapping scenario {and}
propose an $\ell_1$-regularized LS estimation problem to obtain sparse range--angle charts to efficiently facilitate subsequent mapping phases, {while also shortly quantify the involved processing complexity}.
{It is noted that the presentation in this section considers an individual arbitrary sensing location $\mathbf{p}^{\text{UE}}_{l}$. For readers' convenience, the main variables used in this section are summarized in Table~\ref{tab:table_section3}.}

\subsection{Problem Statement}

First, we extend the signal model in \eqref{eq:RX_signal} to include multiple sensing directions. For the $\thh{i}$ sensing direction with azimuth angle $\varphitilde_i$, with $i = 0, \dots, I-1$, and the $\thh{m}$ TX symbol, the frequency-domain vector $\mathbf{x}_{i,m}= [x_{i,m,0}, \dots ,x_{i,m,N-1}]^T  \in \complexset{N}{1}$ is transmitted. The OFDM frequency-domain measurement vector observed and collected by the RX antenna array, containing all the target reflections, can then be expressed as
\begin{align}\label{eq_signal}
    \yy_{i,m} = \sum_{p=0}^{C_{\text{R}}-1} \sum_{q=0}^{\Nphi-1} \bpq \, g(\tilde{\varphi}_q - \varphitilde_i) \, \cc(\tilde{\tau}_p) \odot \xx_{i,m} + \vv_{i,m}, 
\end{align}
where $\odot$ denotes the Hadamard (element-wise) product, 
$C_{\text{R}}$ and $\Nphi$ are the numbers of range and azimuth cells in the discretized range--angle chart, respectively, and $\bpq$ is the reflection coefficient at the $\thh{(p,q)}$ range--angle cell located at $(\tilde{\tau}_p, \tilde{\varphi}_q)$. The variables $g(\varphi)$ and $\vv_{i,m} = [v_{i,m,0}, \dots ,v_{i,m,N-1}]^T  \in \complexset{N}{1}$ refer to the combined TX-RX antenna pattern, defined in (\ref{eq_pattern_g}), and the frequency-domain additive noise vector, respectively. Furthermore, in~(\ref{eq_signal}), the frequency-domain steering vector for a specific cell's delay $\tilde{\tau}_p$ reads 
\begin{align}
    \cc(\tilde{\tau}_p) \triangleq   \left[ 1, e^{-j 2 \pi \deltaf \tilde{\tau}_p}, \ldots,  e^{-j 2 \pi (N-1) \deltaf  \tilde{\tau}_p} \right]^T.
\end{align}

\renewcommand{\arraystretch}{1.25}
\begin{table}[t]
  \begin{center}
  \setlength{\tabcolsep}{4.5pt}
    \caption{{\textsc{Basic Notations and Variables in Range--Angle Processing}}}
    \label{tab:table_section3}
    \begin{tabular}{|c || c | c |}
      \hline
      \textbf{Variable} & \textbf{Definition}  \\
      \hline 
      \hline
      $n$ & OFDM subcarrier index \\
      \hline
      $N$ & Number of active OFDM subcarriers\\
      \hline
      $m$ & OFDM symbol index \\
      \hline
      $M$ & Number of OFDM symbols per beam\\
      \hline
      $i$ & Sensing direction/beam index \\
      \hline
      $I$ & Number of sensing directions/beams\\
      \hline
      $\xx_{i,m}$ & Frequency-domain transmit vector \\
      \hline
      $\yy_{i,m}$ & Frequency-domain receive vector \\
      \hline
      $\vv_{i,m}$ & Frequency-domain additive noise vector \\
      \hline
      $\cbig$ & Frequency-domain steering matrix \\
      \hline
      $\bbig$ & Reflection coefficient matrix \\
      \hline
      $\ggb(\varphi)$ & Combined TX-RX antenna pattern vector \\
      \hline
    \end{tabular}
  \end{center}
\end{table}

The purpose of the range--angle processing is to estimate the reflection coefficients $\bpq = (\bbig)_{p,q}$ at predefined range--angle grid locations $(\tilde{\tau}_p, \tilde{\varphi}_q)$ with $p=0, \hdots, C_{\text{R}}-1$ and $q=0,\hdots,\Nphi-1$, while exploiting the knowledge of the combined antenna pattern. Note that $\bpq = 0$ if there is no scatterer at the $\thh{(p,q)}$ range-azimuth cell. For notational convenience, we can next rewrite the frequency-domain received signal or measurement vector \eqref{eq_signal} as
\begin{align}\nonumber
    \yy_{i,m} &=  \sum_{p=0}^{C_{\text{R}}-1} \bb_p^T \ggb(\varphitilde_i) \, \cc(\tilde{\tau}_p) \odot \xx_{i,m} + \vv_{i,m} \\ \nonumber
    &= \xx_{i,m} \odot \left( \sum_{p=0}^{C_{\text{R}}-1} \cc(\tilde{\tau}_p) \bb_p^T \right) \ggb(\varphitilde_i) + \vv_{i,m} \\ \label{eq_yyi}
    &= \xx_{i,m} \odot \cbig \bbig \ggb(\varphitilde_i) + \vv_{i,m} \, ,
\end{align}
where
\begin{subequations}
\begin{align} \label{eq_cbig}
    \bb_p &\triangleq \left[ b_{p,0}, \, \ldots, \, b_{p,\Nphi-1} \right]^T ~, \\
    \ggb(\varphitilde_i) &\triangleq \left[ g(\tilde{\varphi}_0 - \varphitilde_i), \, \ldots, \, g(\tilde{\varphi}_{\Nphi-1} - \varphitilde_i) \right]^T ~, \\
    \cbig &\triangleq \left[ \cc(\tilde{\tau}_0) \, , \ldots \, , \cc(\tilde{\tau}_{C_{\text{R}}-1}) \right] \in \complexset{N}{C_{\text{R}}} ~, \\
    \bbig &\triangleq \left[ \bb_0 \, , \ldots \, , \bb_{C_{\text{R}}-1} \right]^T \in \complexset{C_{\text{R}}}{\Nphi} \label{eq:rangeAngleChart_B}~.
\end{align}
\end{subequations}
Specifically, the matrix $\bbig$ defined in (\ref{eq:rangeAngleChart_B}) represents the range--angle chart to be estimated.
Aggregating \eqref{eq_yyi} over all $I$ sensing directions, the observation matrix for the $\thh{m}$ OFDM symbol at a given UE location can be expressed as
\begin{align}\label{eq_obs}
    \ybig_m = \xbig_m \odot \cbig \bbig \gbig + \vbig_m ~,
\end{align}
for $m = 0, \ldots, M-1$, where
\begin{subequations}
\begin{align}
     \ybig_m &\triangleq \left[ \yy_{0,m} \, , \ldots \, , \yy_{I-1,m}  \right] \in \complexset{N}{I} ~,\\
     \xbig_m &\triangleq \left[ \xx_{0,m} \, , \ldots \, , \xx_{I-1,m}  \right] \in \complexset{N}{I} ~,\\
     \gbig &\triangleq \left[ \ggb(\varphitilde_0) \, , \ldots \, , \ggb(\varphitilde_{I-1})  \right] \in \complexset{\Nphi}{I} ~,\\
     \vbig_m &\triangleq \left[ \vv_{0,m} \, , \ldots \, , \vv_{I-1,m}  \right] \in \complexset{N}{I} ~.
\end{align}
\end{subequations}

At any given $\thh{l}$ UE location, the range--angle estimation problem can then be formally defined as follows. Given the transmit symbols $\{\xbig_m\}_{m=0}^{M-1}$, the steering matrix $\cbig$ (known through the delay grid locations $\{\tilde{\tau}_p\}_{p=0}^{C_\text{R}-1}$),
the antenna pattern matrix $\gbig$ (known through the angular grid locations $\{\tilde{\varphi}_q\}_{q=0}^{\Nphi-1}$
and the sensing directions $\{\varphitilde_i\}_{i=0}^{I-1}$),
estimate the range--angle chart $\bbig$ from the observation $\{\ybig_m\}_{m=0}^{M-1}$ in \eqref{eq_obs}.

\subsection{Regularized LS for Sparse Range--Angle Processing}\label{sec:RangeAngleProc}
To harness the sparsity of the mm-wave propagation channels, and thus to obtain sparse range--angle charts {from \eqref{eq_obs}}, we {propose to} consider {the following} $\ell_1$-regularized {LS problem:}
\begin{align}\label{eq_ls_cs}
 \min_{\bb} \normsmall{\yy - \boldPhi \bb }_2^2 + \lambda \normsmall{\bb}_1 ~,
\end{align}
{where $\bb \triangleq \vecc{\bbig}$, $\yy = [ \yy_0^T, \, \ldots , \, \yy_{M-1}^T]^T$, $\yy_m \triangleq \vecc{\ybig_m}$, $\boldPhi$ is a dictionary matrix defined in Appendix~\ref{app_ls}, and $\lambda$ denotes the regularization parameter. {Here, 
 $\vecc{\cdot}$ denotes the vectorization operator.}
For reference, Appendix~\ref{app_ls} also describes the standard LS solution $\bbighatls$ of \eqref{eq_ls_cs} without regularization.} 
To solve the sparse map recovery problem in \eqref{eq_ls_cs}, we resort to the iterative shrinkage/thresholding algorithm (ISTA) \cite{daubechies2004iterative,beck2009fast,ISTA_EM_2003}, as outlined in Algorithm~\ref{alg_ista}, where $\lambdamax(\cdot)$ denotes the maximum eigenvalue of a matrix and $(\cdot)_{+}$ yields the positive part of a real number while ${\rm{sgn}}(z) = z / \lvert z \rvert$ yields the sign of a complex number. 

{
The obtained sparse range--angle charts are then subject to a target detection phase that provides the input for the subsequent mapping phases, described in Section~IV. Specifically, a local maximum detection is implemented to the ISTA output, considering a minimum target separation in both range and angle domains. 
In addition, we limit the maximum number of detected targets per sensing location to maintain the complexity of the following tracking-based dynamic mapping phase. Moreover, a threshold test is considered to discard weak and thereon irrelevant target reflections. Further specifics in terms of the numerical threshold values are provided along the numerical results in Sections V and VI.
}

Finally, the corresponding Cartesian coordinates can be calculated for any given point of the range--angle chart as follows. 
Considering a UE location $\mathbf{p}^{\text{UE}}_{l} = \left (x_{l}^{\text{UE}},y_{l}^{\text{UE}}\right )$ with location index $l\in \left [0,L-1\right ]$, as illustrated in Fig.~\ref{fig:ellipseModel}, the Cartesian coordinates of the $\thh{(p,q)}$ range--angle cell located at $(\tilde{\tau}_p, \tilde{\varphi}_q)$ can be calculated as
\begin{equation}
\label{eq:X_Y_coordinates}
\begin{split}
    {x}_{a} & = x_{l}^{\text{UE}} + \rho(\tilde{\tau}_p,\tilde{\varphi}_q+ \varphi^{\text{OR}}_l)\cos(\tilde{\varphi}_q+ \varphi^{\text{OR}}_l), \\
    {y}_{a} & = y_{l}^{\text{UE}} + \rho(\tilde{\tau}_p,\tilde{\varphi}_q+ \varphi^{\text{OR}}_l)\sin(\tilde{\varphi}_q+ \varphi^{\text{OR}}_l) ,\\
\end{split}    
\end{equation}
where $a = \left \{ p + q C_{\text{R}}  + l C_{\text{R}} \Nphi, \forall p, \forall q,\forall l  \right \}$, and $\rho(\tau,\varphi)$ is defined in (\ref{eq_UEdistance}). The parameter $\varphi^{\text{OR}}_l$ denotes the UE orientation angle as illustrated in Fig.~\ref{fig:ellipseModel}.

{\subsection{Complexity Analysis}}
{The computational complexity of Algorithm~\ref{alg_ista} can be analyzed in two stages. First, the initialization stage requires the computation of the LS solution in \eqref{eq_bhat_ls2}, which can be handled very efficiently since the matrices $\left( \cbig^H \cbig\right)^{-1} \cbig^H$ and $\gbig^H \left( \gbig \gbig^H \right)^{-1}$ can be pre-computed offline. In particular, the complexity of the LS solution is given by $\mathcal{O}((MN+C_{\text{R}}N+C_{\text{R}} \Nphi) I)$, which results from $MNI$ multiplications in the coherent integration and the two matrix multiplications with $\mathcal{O}(C_{\text{R}} N I)$ and $\mathcal{O}( C_{\text{R}} I \Nphi)$ operations. Second, for the iterations stage, the coherent integration term has already been computed in \eqref{eq_bhat_ls2} and does not change during the iterations. Similarly, the matrices $\cbig^H \cbig$ and $\gbig \gbig^H$ can be stored offline before executing the iterations. Hence, the per-iteration complexity of Algorithm~\ref{alg_ista} is dictated by the two matrix multiplications involved in $\cbig^H \cbig \bbig^{(k)} \gbig  \gbig^H$, leading to $\mathcal{O}(C_{\text{R}}^2 \Nphi + C_{\text{R}} \Nphi^2)$ operations. To gain further insights, let the total number of range--angle grid points be denoted by $C = C_{\text{R}} \Nphi $ and assume $C_{\text{R}}  = \Nphi = I$ for simplicity. Then, the overall complexity becomes $\mathcal{O}(MNI + CN + N_{\rm{iter}} C^{1.5})$, where $N_{\rm{iter}}$ is the number of iterations. 
In practice, we have observed that the algorithm converges in few tens of iterations for the RF measurement data.
Consequently, the complexity of Algorithm~\ref{alg_ista} is linear in the total number of transmit symbols $MNI$ and sub-quadratic in the number of grid points $C$.}

\begin{algorithm}[t]
	\caption{Iterative Shrinkage/Thresholding Algorithm (ISTA) {\cite{daubechies2004iterative,beck2009fast,ISTA_EM_2003}} for Sparse Range--Angle Processing in \eqref{eq_ls_cs}}
	\label{alg_ista}
	\begin{algorithmic}
	    \State \textbf{Input:} Observation $\yy$ in \eqref{eq_y_all}, step size $\eta$, regularization parameter $\lambda$.
	    \State \textbf{Output:} Range--angle chart $\bbighatista$.
	    \State \textbf{Initialization:} Set $k=0$, $\bbig^{(0)} = \bbighatls$ and ${\eta = \frac{\beta}{ \lambdamax(\cbig^H \cbig) \lambdamax(\gbig \gbig^H) }}$ for some $\beta \in (0, \, 1)$.
	    \State \textbf{Repeat} 
	    \Indent 
	    \State \vspace{-0.25in}
	    \begin{align}
	        &\bbig^{(k+1)} = \tlambda{\lambda \eta} \bigg( \bbig^{(k)} - 2 \eta \cbig^H \bigg[  \cbig \bbig^{(k)} \gbig  \\ \nonumber &~~~~~~~~~~~~~~~~ - \frac{1}{M} \sum_{m=0}^{M-1}( \xbig_m^{*} \odot \ybig_m ) \bigg] \gbig^H \bigg)~,
	        \\ \nonumber &k = k + 1 ~,
	    \end{align}
	    where 
	    \begin{align}
	        \tlambda{\alpha}(\bbig)_{r,q} = (\lvert B_{r,q} \rvert - \alpha)_{+} {\rm{sgn}}( B_{r,q})~.
	    \end{align}
	    \EndIndent
	    \State \textbf{until convergence}
	\end{algorithmic}
	\normalsize
\end{algorithm}

\vspace{3mm}
\section{Tracking-based Dynamic Mapping }
\label{sec:trackingBasedDynamicMapping}

In this section, we address the actual mapping task based on a sequence of sensing measurements (range-angle charts and the corresponding identified targets) obtained by a moving UE at locations $\mathbf{p}^{\text{UE}}_{l}$, $l = 0,\dots, L-1$. Specifically, we propose and formulate a novel state model for mapping both diffuse scattering and specular reflections, which allows efficient tracking of individual scatterers or reflecting surfaces over time using IMM {extended Kalman filter} (EKF), {where the ISTA outputs are considered as measurements}. To remove second-order reflections or other second-order channel interactions, a novel measurement selection process is also presented. 
{The tracking-based mapping allows to efficiently utilize the correlations in the sequences of measurements obtained by moving sensors, while also enables tracking of the possible changes in the radio environment. }

{\subsection{Fundamentals and Rationale}~\label{sec:dynamicMappingIntro}}
{At each sensing instant, the environment can be represented as a collection of discrete scatterers, which correspond to the physical points of interaction of the radio waves with the environment. These include specular reflections of the signal off the walls and other flat surfaces
, and diffuse scattering off the multifaceted objects or objects with rough surfaces \cite{Pinel2013}. Tracking time evolution of the interaction points' positions can help to reconstruct the geometry of the considered environment \cite{Oren2002}. To this end, we consider a sequence of known UE sensing positions denoted as $\mathbf{p}^{\text{UE}}_{l} = \left (x_{l}^{\text{UE}},y_{l}^{\text{UE}}\right )$, $l = 0,\dots, L-1$, where at position $l$ the $K_l$ points of interaction with absolute 2D coordinates are denoted by $\mathbf{p}^{\text{SC}}_{l,k} = ({x}^{\text{SC}}_{l,k},{y}^{\text{SC}}_{l,k})$, $k = 1,\dots, K_l$. 
Specular reflections and diffuse scattering lead to different evolutions of these interaction points:
\begin{itemize}
    \item \emph{Specular reflections:} there is an unambiguous relationship between the incidence and reflection angles,so that  the location of specular reflection point at each time $l$ is a function of the positions of TX and RX at the UE, while the motion of the UE causes the shift of the specular interaction point. In this work, we consider the continuous white noise acceleration (CWNA) motion model \cite{BarShalom2001}, implying that the interaction point moves with constant velocity perturbed with a white noise process, to represent the kinematics of the moving scatterers. 
    \item \emph{Diffuse scattering:} a relatively wide distribution of the reflection angles may correspond to each incident angle. Due to this dispersion of the reflected wave, the apparent location of the interaction point remains constant for several consecutive UE locations. To describe the kinematics of these stationary scatterers, we consider scatterer position to be constant, perturbed with a white noise process. In accordance with naming convention of~\cite[Ch.~6]{BarShalom2001}, we call this approach a continuous white noise velocity (CWNV) model in the continuation.  
\end{itemize}
}

{However, the above classification is still incomplete as 
the indoor propagation environment may give rise to multi-bounce paths, which have different apparent mobility. To filter out such paths, we thus first present a measurement selection process, after which the specular reflections and scattering points will be tracked using an IMM EKF.}
{Furthermore, the reflections from concave corners, albeit they can be formally classified as double specular reflections, are useful for mapping the environment. Their kinematics can be described by the CWNV model.}
{We note that terminology-wise, the range and angle estimates obtained through the ISTA-based range-angle charting serve as the main inputs or measurements for the IMM EKF, while for the measurement selection purposes we also assume that the corresponding received signal strength (RSS) measurements are available. For readers' convenience, the basic measurement selection and IMM EKF notations are summarized in Table~II.} {Specifically, we denote the range and angle estimates obtained through range-angle charting by $\hat{d}^{\text{UE}}_{l,k}$ and $\hat{\varphi}^{\text{UE}}_{l,k}$, respectively.}

\renewcommand{\arraystretch}{1.25}
\begin{table}[t]
  \begin{center}
  \setlength{\tabcolsep}{4.5pt}
    \caption{{\textsc{Main Notations and Variables in Measurement Selection and Tracking-based Dynamic Mapping}}}
    \label{tab:table_section4}
    \begin{tabular}{|c || c | c |}
      \hline
      \textbf{Variable} & \textbf{Definition}  \\
      \hline 
      \hline
      $\mathbf{p}^{\text{SC}}_{l,k}$ & Tracked scatterer location \\
      \hline
      {$\hat{\varphi}^{\text{UE}}_{l,k},\hat{d}^{\text{UE}}_{l,k}$}  & Target angle  and distance measurements \\
      \hline
      $\text{{RSS}}^{\text{UE}}_{l,k}$ & Target RSS measurement \\
      \hline
      ${\mathcal{H}_{0}, \mathcal{H}_{1}}$ & Measurement selection hypotheses \\
      \hline
      $\alpha^\prime, \beta, \beta^\prime$ & Parameters of measurement selection models \\
      \hline
      $p^{\text{prior}}, p^{\text{post}}$ & Prior and posterior sampling probability estimate \\
      \hline
      $d_\textrm{th}$ & Threshold distance \\
      \hline
      ${\mathsf{M}^j}$ & IMM (sub)model $j$ \\
      \hline
      $\mu^j_0$ & Prior probability of a (sub)model $j$ \\
      \hline
       $\mathbf{s}^{\mathsf{M}_j}_{l,k}$ & State vector corresponding to model $\mathsf{M}_j$ \\
      \hline
      $\mathbf{C}_{l,k}^{\mathsf{M}_j}$ & State covariance matrix corresponding to model $\mathsf{M}_j$ \\
      \hline
        $\mathbf{F}^{\mathsf{M}_j}$ & State-transition matrix corresponding to model $\mathsf{M}_j$\\
        \hline
        $\mathbf{Q}^{\mathsf{M}_j}$ & Process noise covariance matrix for model $\mathsf{M}_j$\\
         \hline
        $\mathbf{R}$ & Measurement covariance matrix\\
      \hline
        $\mathbf{m}_{l,k}$ & Measurement vector \\
     \hline
      $\mathbf{h}(.)$ & Observation function \\
      \hline
      $\mathbf{H}_{l,k}$ & Jacobian matrix \\
      \hline
      $D^{a}_{k,q}$ & Measurement association distance \\
      \hline
      
      \hline
    \end{tabular}
  \end{center}
\end{table}

\subsection{Tracking Filter Measurement Selection Method}\label{sec:measurementSelectio}
{As noted above, the overall available mapping-related measurements for the $\thh{k}$ scatterer obtained at the $\thh{l}$ UE location $\mathbf{p}^{\text{UE}}_{l}$ are
{$[\hat{\varphi}^{\text{UE}}_{l,k}, \hat{d}^{\text{UE}}_{l,k}, \text{\small{RSS}}^{\text{UE}}_{l,k}]_{k=1}^{K_l}$,
where $\hat{\varphi}^{\text{UE}}_{l,k}$ and $\hat{d}^{\text{UE}}_{l,k}$} are the angle and distance estimates, respectively, obtained through range-angle charting as described in Section~\ref{sec:rangeAngle_processing}, while $\text{\small{RSS}}^{\text{UE}}_{l,k}$ is the corresponding RSS measurement.}

\begin{figure}[t]
        \centering
        \includegraphics[trim=1.5cm 1.5cm 2cm 2cm,clip,width=1\columnwidth]{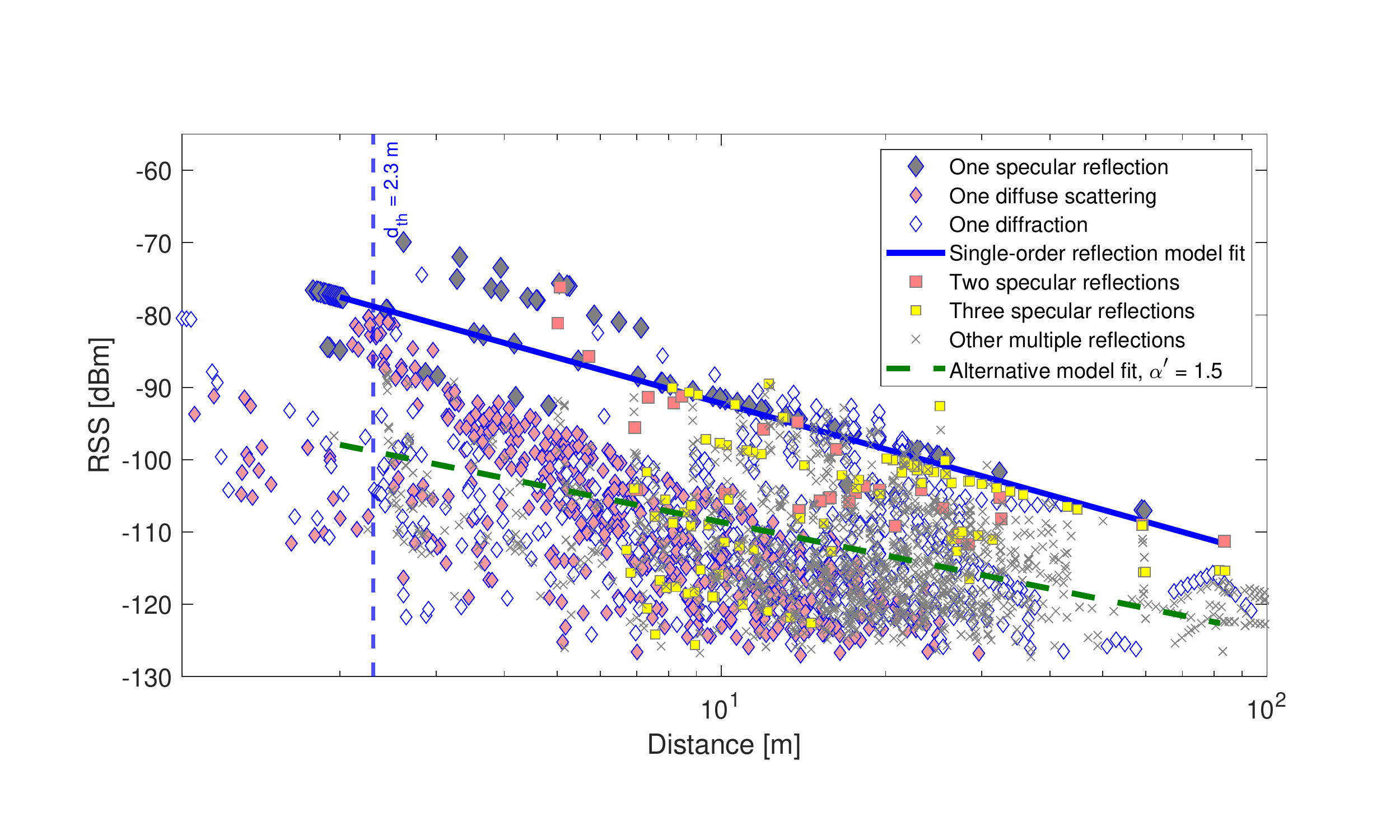}
        \vspace{-0.5cm}
        \caption{ 
        {Illustration of the measurement selection approach building on two alternative pathloss models in (\ref{eq:freespace}) with ray-tracing data from the indoor scenario described in Section V. Distance threshold $d_\textrm{th}$ is also shown.
        } 
        } 
        \label{fig:measurement_selection}
\end{figure}

{From the mapping point of view, the measurements originating from the single interactions are of a particular interest, hence we will harness the RSS to distinguish the first-order reflections from the higher-order interactions.} 
In particular, we consider two alternative path loss model hypotheses, {expressed in logarithmic scale}: $\mathcal{H}_0$ corresponds to single-order reflection model and $\mathcal{H}_1$ corresponds to higher-order reflection model, with 
\begin{align}\label{eq:freespace}
    \text{\small{RSS}}^{\text{UE}}_{l,k} = 
    \begin{cases}
    {\beta - 20 \log_{10}(d^{\text{UE}}_{l,k}) + n^{(0)}_{l,k}}  & \mathcal{H}_0 \\
    {\beta^\prime - 10\alpha^\prime \log_{10}(d^{\text{UE}}_{l,k}) + n^{(1)}_{l,k}} &  \mathcal{H}_1
    \end{cases}
\end{align}
where $\beta$ and $\beta^\prime$ are power scaling parameters depending, e.g., on transmit power, carrier frequency and reflection coefficient, $\alpha^\prime$ is the path loss exponent (set to $2$ under $\mathcal{H}_0$), and $n^{(i)}_{l,k} \sim \mathcal{N}(0,\sigma_{\mathcal{H}_i}^2)$, $i \in \{0,1\}$ is RSS measurement noise, where $\sigma_{\mathcal{H}_i}^2$ is a design parameter. In Fig.~\ref{fig:measurement_selection}, the RSS measurements $\text{\small{RSS}}^{\text{UE}}_{l,k}$ are illustrated {as a function of distance $d^{\text{UE}}_{l,k}$} for all measurements of the ray-tracing data from the indoor scenario described in Section \ref{sec:Implementation}. It can be seen that only a relatively small fraction of all measurements are actually single reflections and that at short distances, say {$d^{\text{UE}}_{l,k} \le d_\textrm{th}$}, where  $d_\textrm{th}$ is on the order of 2-3 meters, only single-order scattering occurs.  

To estimate the unknown parameters $[\beta, \beta^\prime, \alpha^\prime]$ and the classification of each measurement, denoted by $\zeta_{l,k}\in \{\mathcal{H}_0,\mathcal{H}_1\}$,  we use the expectation maximization (EM) algorithm \cite{Kay93}.  
The measurement selection process is performed recursively over different measurement locations, using the pathloss parameter estimates from previous time steps as initial estimates for the following time step. We set the prior $p^{\text{prior}}({\zeta}_{l,k} =\mathcal{H}_0)=0.5$. 
As a result of the expectation step of the EM algorithm, each RSS measurement is associated with a posterior probability estimate $p^{\text{post}}({\zeta}_{l,k} = \mathcal{H}_0) \in [0,1]$, which indicates the probability that the measurement is originated from the single-interaction model. The final criterion for measurement selection of first-order reflection or scattering is given as
\begin{equation}
\label{eq:measurement_selection}
p^{\text{post}}({\zeta}_{l,k} = \mathcal{H}_0) \geq p^\textrm{sampl}_\textrm{th} \text{  or  }
{\hat{d}^{\text{UE}}_{l,k} \leq  d_\textrm{th}},
\end{equation}
where $p^\textrm{sampl}_\textrm{th}$ and $d_\textrm{th}$ are design parameters, which affect the sensitivity of the measurement selection process. 

{\subsection{IMM EKF Tracking and Measurement Association}\label{sec:IMM_EKF}}
{In this sub-section, we present the IMM EKF based tracking of the scatterer locations. Each of the $K$ scatterers is in general tracked separately, but we omit the scatterer index below for notational simplicity. 

\subsubsection{IMM models and weights}
For generality, let us assume that there exist a set of (sub)models $\{\mathsf{M}_1,...,\mathsf{M}_W\}$, and a set of corresponding prior probabilities {$\mu^j_0 = P\{\mathsf{M}^j_0\}, j = 1,...,W$.} 
At each time instant, the probability for each scatterer to switch from model $i$ to model  $j$  are assumed known and denoted by $p_{ij} = P\{\mathsf{M}^j_l|\mathsf{M}^i_{l-1}\}, i,j = 1,..,W$. Both probability sets are IMM filter design parameters and are chosen to reflect the properties of the environment. 
At each time step, one obtains the initial conditions for each model 
using all previous state estimates. 
Also the probabilities of each model are updated at every time step. The IMM combined state estimate and the covariance are then calculated as a weighted mean of a-posteriori EKF state estimates and the covariance matrices of all the models, calculated via the standard EKF prediction and update \cite{HSS11}, where the weights are defined by the probabilities.} {The details of the IMM processing steps and the propagation of the model probabilities can be found in \cite{Nadarajah2012}.}

{In our case, $W = 2$, and as the corresponding IMM sub-models $\{\mathsf{M}_1,\mathsf{M}_2\}$ we consider two EKFs utilizing CWNV or CWNA motion models for the state dynamics. {Note that prior model probabilities $\mu^j_0$ depend on the expected proportion between specular reflections and diffuse scattering and may generally differ for different environments and carrier frequencies.} 
The EKF state vector at time instant $l$, corresponding to the UE location $\mathbf{p}^{\text{UE}}_{l}$, is generally denoted as $\mathbf{s}_l$. {Then, depending on the considered motion model, the state vector includes either 2D position of a scatterer (the CWNV model)  
for tracking diffuse scattering points, or scatterer's position and velocity (the CWNA model) 
for tracking specular reflections, expressed respectively as 
\begin{align}
\begin{split}
\mathbf{s}^{\mathsf{M}_1}_{l,k} & = [x^{\mathsf{M}_1}_{l,k},y^{\mathsf{M}_1}_{l,k}]^T,\\
\mathbf{s}^{\mathsf{M}_2}_{l,k} & = [x^{\mathsf{M}_2}_{l,k},y^{\mathsf{M}_2}_{l,k},\dot{x}^{\mathsf{M}_2}_{l,k},\dot{y}^{\mathsf{M}_2}_{l,k}]^T.
\end{split}
\end{align}
}
The state vector for each model is propagated following the standard linear EKF state transition model with the non-linear measurement model 
{\cite{Kay93,HSS11}. 
This can be expressed as
\begin{align} \label{eq:ekf_prediction}
\begin{split}
\mathbf{s}^{\mathsf{M}_j}_{l,k} & =  \mathbf{F}^{\mathsf{M}_j} \mathbf{s}^{\mathsf{M}_j}_{l-1,k} + \mathbf{u}^{\mathsf{M}_j}_{l,k} , \\
\mathbf{m}_{l,k} & = \mathbf{h}(\mathbf{\bar{s}}^{\mathsf{M}_j}_{l,k}) + \mathbf{e}_{l,k}.
\end{split}
\end{align}
where $\mathbf{\bar{s}}^{\mathsf{M}_j}_{l,k}=[x^{\mathsf{M}_j}_{l,k},y^{\mathsf{M}_j}_{l,k}]^T$, $\mathbf{F}^{\mathsf{M}_j}$ is the state-transition matrix, $\mathbf{u}^{\mathsf{M}_j}_{l,k} \sim \mathcal{N} \left (0,\mathbf{Q}^{\mathsf{M}_j} \right )$ denotes the state-process noise, and $\mathbf{e}_{l,k} \sim \mathcal{N} \left (0,\mathbf{R} \right )$ is the measurement noise with covariance $\mathbf{R} \in \mathbb{R}^{2\times2}$. 
Expressions for $\mathbf{F}^{\mathsf{M}_j}$ and state-process noise covariance $\mathbf{Q}^{\mathsf{M}_j}$ for both models can be obtained from the discretization of the continuous state-transition model \cite[Ch.~2]{HSS11}.
The related power spectral density $Q^{\mathsf{M}_j}_c$ of the state-process noise is again one of the EKF design parameters. 
We note that the sizes of the state-transition and covariance matrices 
are defined by the used motion model, i.e.,  
$\mathbf{C}_{l,k}^{\mathsf{M}_1}, 
\mathbf{F}^{\mathsf{M}_1},\mathbf{Q}^{\mathsf{M}_1} 
\in \mathbb{R}^{2\times2}$ and 
$\mathbf{C}_{l,k}^{\mathsf{M}_2}, 
\mathbf{F}^{\mathsf{M}_2}, \mathbf{Q}^{\mathsf{M}_2} 
\in \mathbb{R}^{4\times4}$. }

The corresponding update step is more involved due to the unknown association between scatterer positions at one time and the next. We first describe the measurement model and corresponding Jacobians needed in the EKF. Then we detail the solution to the data association problem. 
}

{\subsubsection{Measurement model and Jacobian}
The fundamental radar-based {measurement vector $ \mathbf{m}_{l,k} = [\hat{\varphi}^{\text{UE}}_{l,k},\hat{d}^{\text{UE}}_{l,k}]^T$} for the $\thh{k}$ scatterer obtained at the $\thh{l}$ UE location $\mathbf{p}^{\text{UE}}_{l}$ 
contains the angle and distance {estimates after the ISTA target detection step (see Section~\ref{sec:rangeAngle_processing}). 
Measurement covariance $\mathbf{R}$ 
is in general a function of the accuracy of the angle and range measurements, while the measurement function is problem-specific and defined by the type of the sensing equipment and the techniques used. In this work, measurement errors stem from ISTA-based range-angle charting and the related target selection process, while their distributions in the range and angle domains can be approximated with normal distribution. } 

Furthermore, stemming from (\ref{eq_UEdistance}) and (\ref{eq_TXRXangle}), the relation of {the noiseless} measurements $[\varphi^{\text{UE}}_{l,k},d^{\text{UE}}_{l,k}]^T$ to the tracked scatterer locations $\mathbf{p}^{\text{SC}}_{l,k}$ can be expressed
via the non-linear differentiable observation function as
\begin{align} \label{eq:measurement}
\renewcommand{\arraystretch}{1.5}
\begin{bmatrix} \varphi^{\text{UE}}_{l,k} \\ d^{\text{UE}}_{l,k} \end{bmatrix} = \mathbf{h}(\mathbf{p}^{\text{SC}}_{l,k}) = \begin{bmatrix} \operatorname{atan2} \left (y^{\text{SC}}_{l,k}-y^{\text{UE}}_{l},x^{\text{SC}}_{l,k}-x^{\text{UE}}_l \right ) \\ \frac{\Vert \mathbf{p}^{\text{TX}}_l-\mathbf{p}^{\text{SC}}_{l,k} \Vert + \Vert \mathbf{p}^{\text{RX}}_l-\mathbf{p}^{\text{SC}}_{l,k} \Vert}{2} \end{bmatrix}.
\end{align}
The Jacobian matrix of the observation function, evaluated at the a-priori estimation of the scatterer location $\hat{\mathbf{p}}^{\text{SC}}_{l,k}$, and the measurement residual  $\mathbf{r}_{l,k} =  \mathbf{m}_{l,k} - \mathbf{h}(\hat{\mathbf{p}}^{\text{SC}}_{l,k})$ can be shown to read
\begin{align} \label{eq:measurement_1}
\renewcommand{\arraystretch}{1.5}
\mathbf{H}_{l,k} = \left[ \begin{array}{cc}
-\frac{y^{\text{SC}}_{l,k}-y^{\text{UE}}_{l}}{{d^{\text{UE}}_{l,k}}^2} & \frac{x^{\text{SC}}_{l,k}-x^{\text{UE}}_l}{{d^{\text{UE}}_{l,k}}^2}\\
\frac{x^{\text{SC}}_{l,k}-x^{\text{TX}}_{l}}{d^{\text{TX}}_{l,k}}+\frac{x^{\text{SC}}_{l,k}-x^{\text{RX}}_{l}}{d^{\text{RX}}_{l,k}} & \frac{y^{\text{SC}}_{l,k}-y^{\text{TX}}_{l}}{d^{\text{TX}}_{l,k}}+\frac{y^{\text{SC}}_{l,k}-y^{\text{RX}}_{l}}{d^{\text{RX}}_{l,k}}  \end{array} \right],
\end{align}
where $d^{\text{TX}}_{l,k} = \Vert\mathbf{p}^{\text{TX}}_l-\mathbf{p}^{\text{SC}}_{l,k}\Vert$ and $d^{\text{RX}}_{l,k} = \Vert\mathbf{p}^{\text{RX}}_l-\mathbf{p}^{\text{SC}}_{l,k}\Vert$.
}

\subsubsection{Measurement association}\label{CrossID}
{The association of scatterers observed at two consecutive time instants is unknown and must be inferred during processing. Due to the measurement noise, complicated system geometry and {occasional miss-classified double interactions}, solving this association problem is challenging. 
We consider relatively slow motion of the radar or frequent measurement updates so that an apparent position of a reflection point is not expected to change dramatically during one time step, allowing us to use a proximity argument  to solve this problem.}
{Specifically, assume that $\{ \hat{\mathbf{p}}_{l-1|l,k} \}, k = 1,...,K_{l-1}$, are the IMM predictions of the positions at time instant $l$ for all scatterers that have been identified at time instant $l-1$.} Furthermore, assume that the coordinates of $K_{l}$ scatterers have been coarsely estimated from the observed angles and distances at a time instant $l$,  using the method described in Section~\ref{sec:rangeAngle_processing}, as 
\begin{align}
\tilde{\mathbf{p}}_{l,k} = 
{\left ( \hat{d}^{\text{UE}}_{l,k_{l}} \cos(\hat{\varphi}^{\text{UE}}_{l,k_{l}}), \hat{d}^{\text{UE}}_{l,k_{l}} \sin(\hat{\varphi}^{\text{UE}}_{l,k_{l}}) \right )}. 
\end{align}
To then associate the newly measured scatterers at step $l$ with those tracked at the previous step $l-1$ and to decide whether to continue or discontinue tracking, an association metric is calculated for each pair based on Mahalanobis distance as
{
\begin{align}
\begin{split}
    D^{a}_{k,q} = &\sqrt{(\tilde{\mathbf{p}}_{l,k} - \hat{\mathbf{p}}_{l-1|l,q})^T\hat{\mathbf{C}}_{l,k}^{-1}(\tilde{\mathbf{p}}_{l,k} - \hat{\mathbf{p}}_{l-1|l,q})} \\ &\text{with } k = 1,...,K_{l}, q = 1,...,K_{l-1}.
\end{split}
\end{align}
}
These association distances are then used as input to a linear assignment problem to find the best global association.  
At the second step in scatterer association process, for each pair
{$(\tilde{\mathbf{p}}_{l,k},\hat{\mathbf{p}}_{l-1|l,q})$} under the best association, we check if  $\tilde{\mathbf{p}}_{l,k}$ is within the ${p}$-percent confidence ellipse of IMM weighted position {prediction $\hat{\mathbf{p}}_{l-1|l,q}$} with covariance $\hat{\mathbf{C}}_{l,k}$.  
{We continue the tracking process only for associated scatterers that fulfill this condition and all others are dropped.} For the newly identified scatterers (i.e., not matched to a previously detected scatter point),  
we initialize the IMM EKF filters. For missed scatterers (i.e., not matched to any  $\tilde{\mathbf{p}}_{l,k}$), we use the prior position estimate for the next time instant.

\begin{figure}[t!]
        \centering
        \includegraphics[width=0.9\columnwidth]{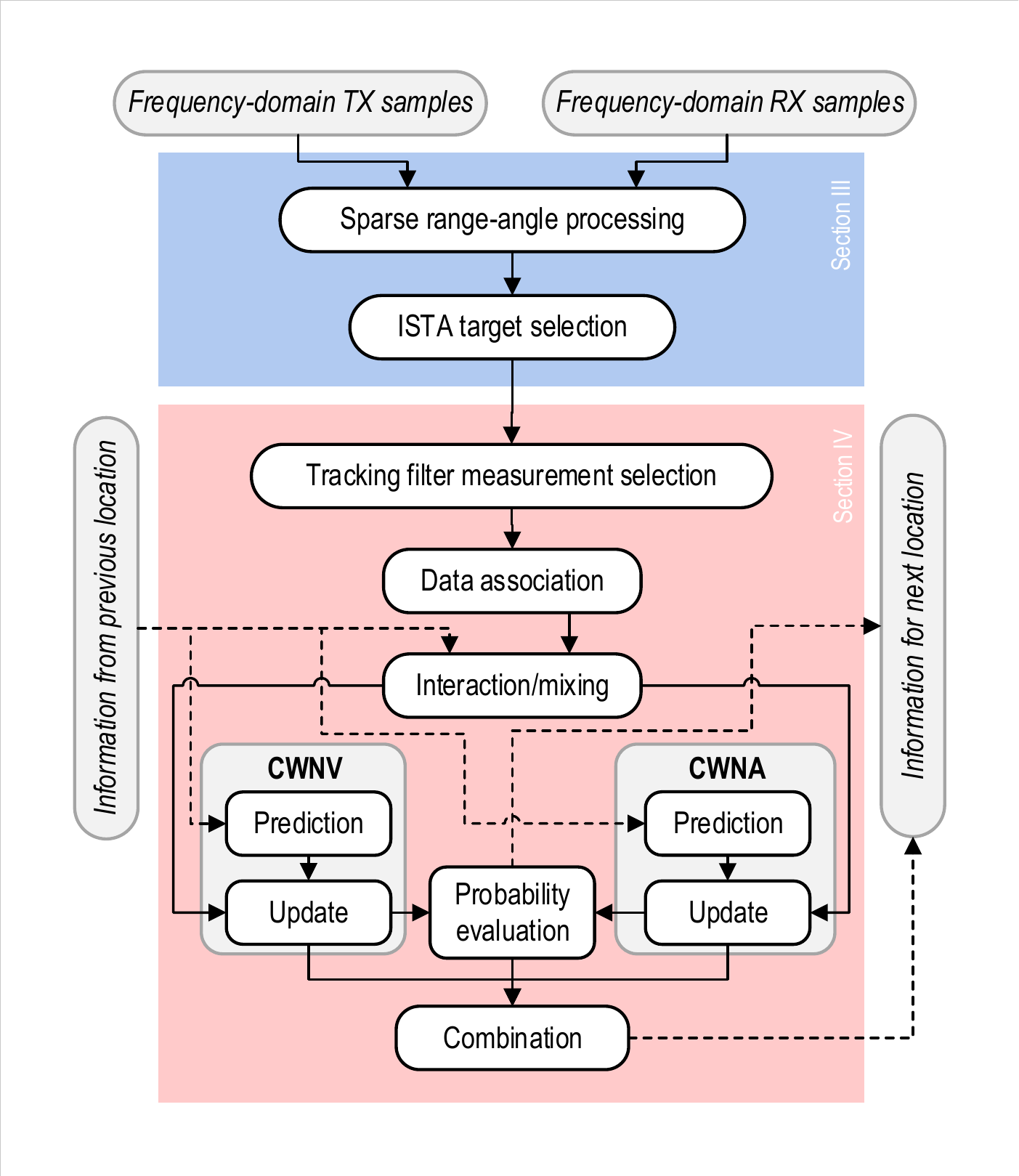}
        \vspace{-3mm}
        \caption{{Overall flow-chart including ISTA range-angle charting, ISTA target selection, measurement selection, data association, and IMM EKF filtering.}
        }
        \label{fig:flowChart}
\end{figure}

{For readers' convenience, an overall flow-chart illustrating and summarizing all the processing steps from the physical-layer IQ signals to the IMM EKF output is shown in Fig.~\ref{fig:flowChart}.}

\vspace{-3mm}
\subsection{{IMM Smoothing}}\label{sec:IMM_smoothing}
In order to utilize all the available radar-based measurements for building a map of the underlying environment, we consider an optional smoothing step. From a variety of available IMM smoothing algorithms we have chosen the approach proposed by \cite{Nadarajah2012}, tailored to our problem, as it is relatively straightforward and does not require storing of the radar distance and angle measurements. In this approach, the IMM smoothing mimics the behaviour of the forward IMM propagation, with each of the interacting IMM sub-models using the so-called Rauch-Tung-Striebel smoothing recursion 
\cite{RTS1965} combined with the model interaction. This becomes possible due to approximation of the backwards model transition probabilities by the forward model transition probabilities, stemming from the Markov properties of the model transition, as well as approximation of the smoothed probability density with the Gaussian mixture of $W$ model conditioned smoothing densities. Such a  mixture can be considered a sufficient statistics of the measurements. 

As the very last stage, a thresholding operation is imposed on the refined covariances of the smoothed scatterer position estimates in order to remove the unreliable estimates. 
We note that  the covariance of the scatterer's position is related to the SNR and to the accuracy of the angle and distance measurements. Hence, this approach allows us to use all available measurement information, automatically discarding the unreliable position estimates. This will be concretely illustrated in Section VI through the processing of both ray-tracing based and the actual RF measurement data.

\section{Evaluation Scenario and Environments}
\label{sec:Implementation}

In this section, the evaluation scenario, ray-tracing environment as well as the experimental RF measurement environment and equipment, used for the validation of the proposed sensing and mapping algorithms are described. {It is also noted that the complete I/Q measurement data is openly available at {\color{blue}\url{https://doi.org/10.5281/zenodo.4475160}} \cite{datasetZenodo_IQdata}.}

\vspace{-3mm}
\subsection{Scenario Description}
The evaluation scenario is an indoor office environment at the Hervanta Campus of Tampere University, Finland, as shown in Fig.~\ref{fig:setup}\subref{fig:setup_scenario}. The considered environment consist of a corridor of 2~m wide and 60~m long with different office rooms on both sides as illustrated by the line art overlaid to all the following mapping results.

{The environment is sensed along straight trajectories, one of which being shown in Fig.~\ref{fig:setup}\subref{fig:setup_scenario}. The sensing related measurements are conducted with distance step of 0.5~m.}
In addition, these positions are deployed in both directions along the corridor, providing a total of two sets of measurements. 
In Fig.~\ref{fig:setup}\subref{fig:setup_scenario}, the considered measurement locations as well as the most significant targets from the radar perspective are shown. We can highlight three walls of adjacent corridors that are perpendicular to the system trajectory -- marked with \textit{A}, \textit{B} and \textit{C} -- located at the left side of the figure. Moreover, at the right side, three metal lockers -- marked with \textit{D}, \textit{E} and \textit{F} -- are expected to be the main targets due to their notable RCS.

\subsection{5G NR Waveform}
In both the RF measurements and the ray-tracing simulations, OFDM-based NR uplink waveform is used with the widest available channel bandwidth at the mm-wave frequency range according to \cite{3GPPTS38104}. Therefore, a channel bandwidth of 400~MHz with subcarrier spacing of 120~kHz is utilized. In particular, for each sensing location and scanning direction, we consider an uplink NR frequency-domain resource grid with $N=3168$ active subcarriers and $M=28$ OFDM symbols corresponding to an observation window of around 0.25~ms. In this case, the $M$ consecutive OFDM symbols are coherently combined to improve the SNR of the obtained range--angle charts as described in (\ref{eq_bhat_ls2}).
According to \cite{myPaper_5_ICC20}, the considered transmit waveform provides a basic radar range resolution of about 40~cm {which is effectively improved by the angular measurements}. 

It is also shortly noted that some recent studies, such as \cite{journal_20,journal_22,journal_28,collab_sahan_TWC2021}, have raised the idea of joint waveform optimization to improve the sensing performance in JCAS type of systems. In this work, however, we deliberately use 3GPP 5G NR standard compliant uplink waveform with all physical channels and signal structures involved, to reflect the true waveform of NR UEs as accurately as possible.

\vspace{-2mm}
\subsection{Ray-Tracing Environment}
\begin{figure}[!t]
    \centering
       \subfloat[Evaluation scenario ]{\includegraphics[width=0.9\columnwidth]{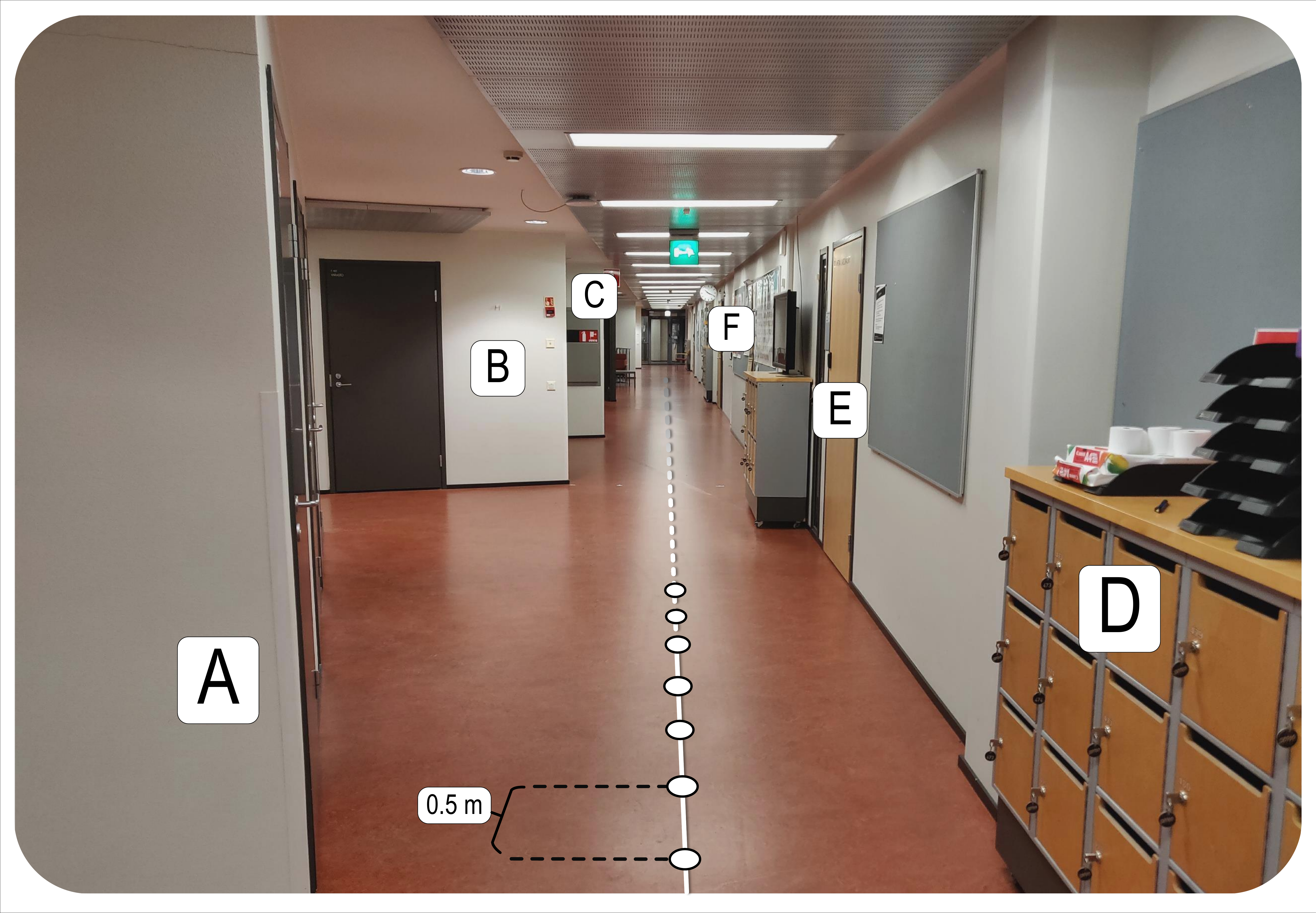}
        \label{fig:setup_scenario}}
        \hfil
        \subfloat[RF measurement equipment]{\includegraphics[width=0.9\columnwidth]{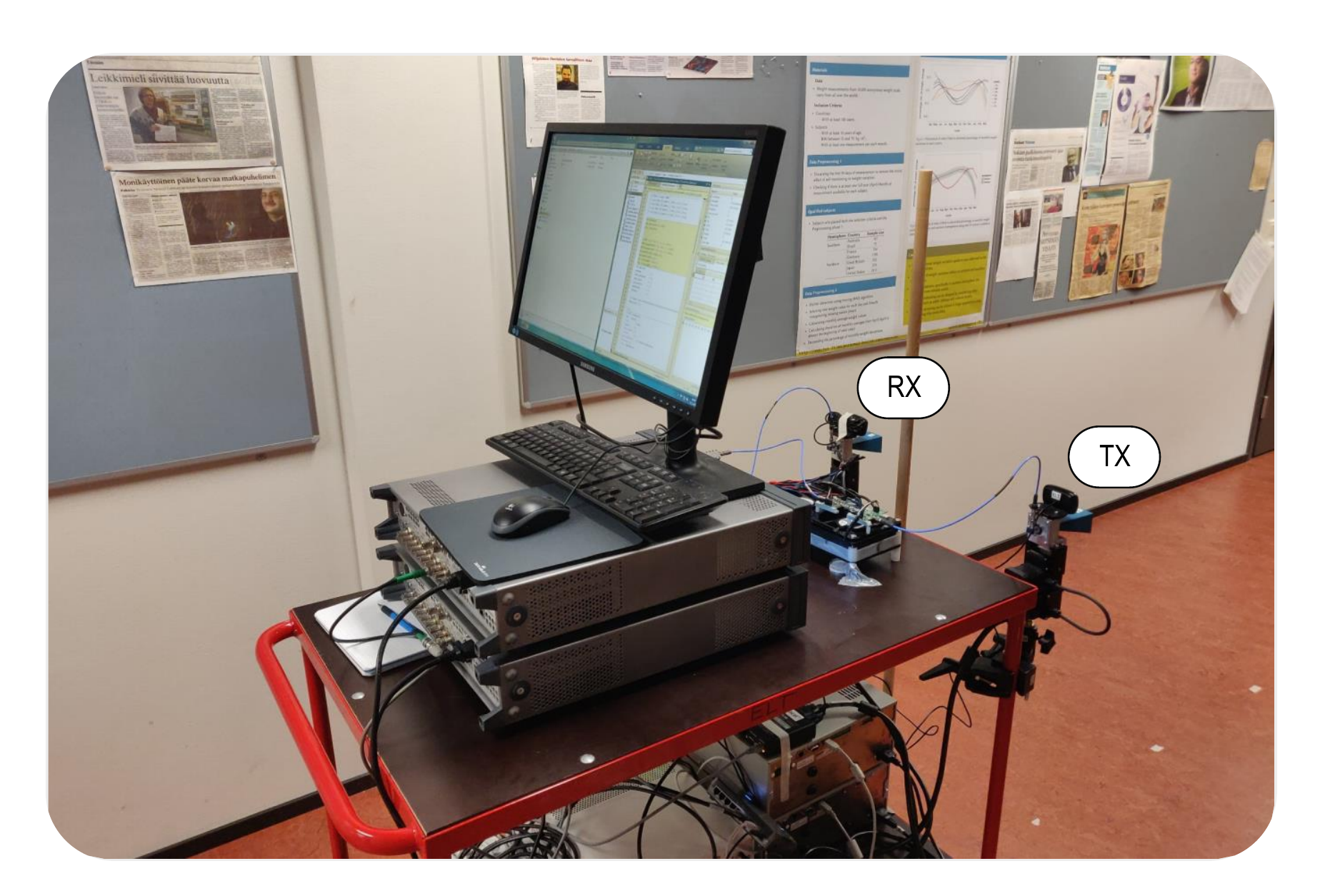}
        \label{fig:setup_equipment}}
    \caption{(a) The indoor evaluation scenario including the main sensing locations and the most important radar targets. (b) The main equipment used in the actual RF measurements at 28~GHz.}
    \label{fig:setup}
\end{figure}

\begin{figure}[!t]
    \centering
        \subfloat[{ISTA-based static mapping} ]{\includegraphics[width=1\columnwidth]{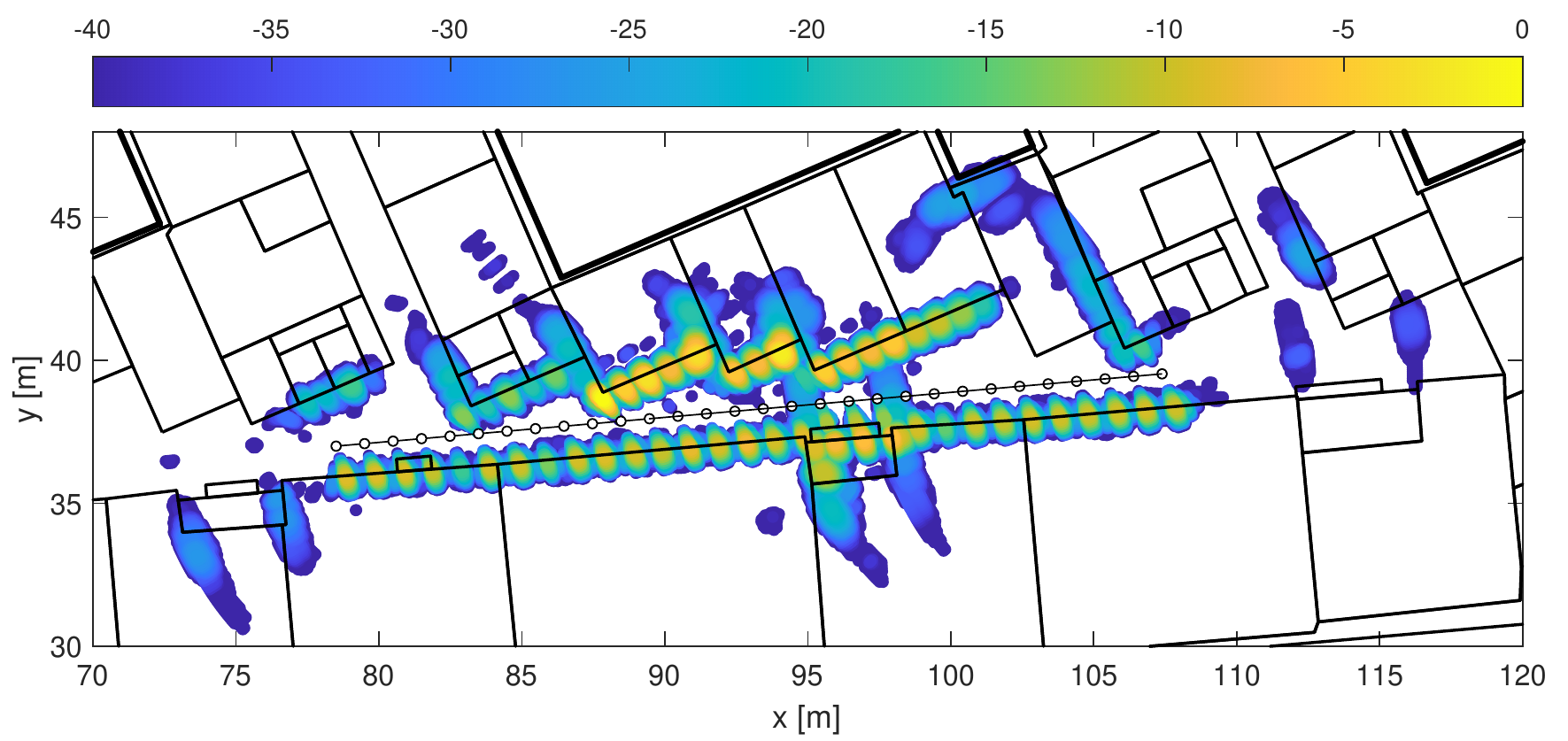}
        \label{fig:mapping_RT_ISTA}}
        \hfill
        \subfloat[Grid-based static mapping \cite{myPaper_4_ICL20} ]{\includegraphics[width=1\columnwidth]{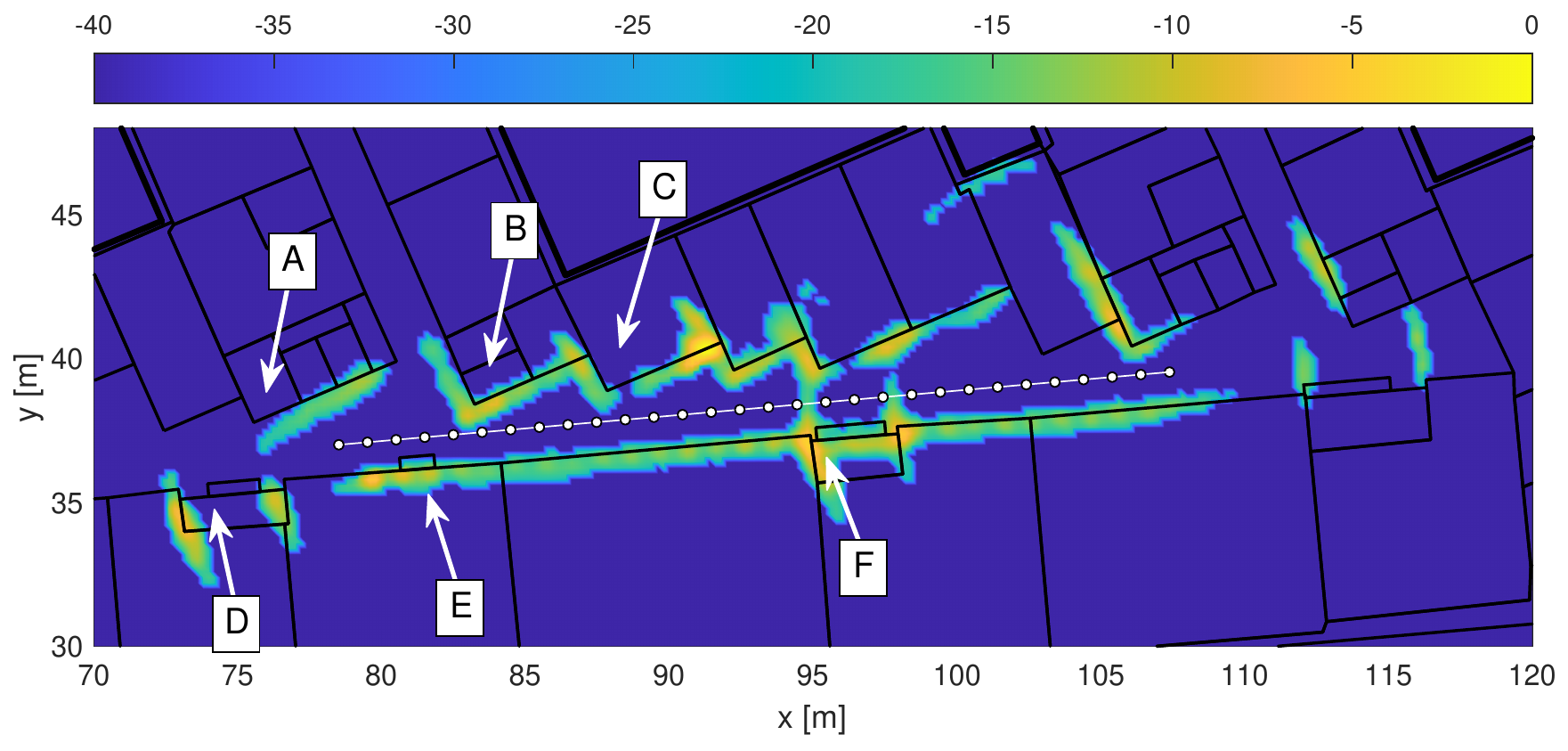}
        \label{fig:mapping_RT_grid}}
        \hfil
        \subfloat[Tracking-based dynamic mapping]{\includegraphics[width=1\columnwidth]{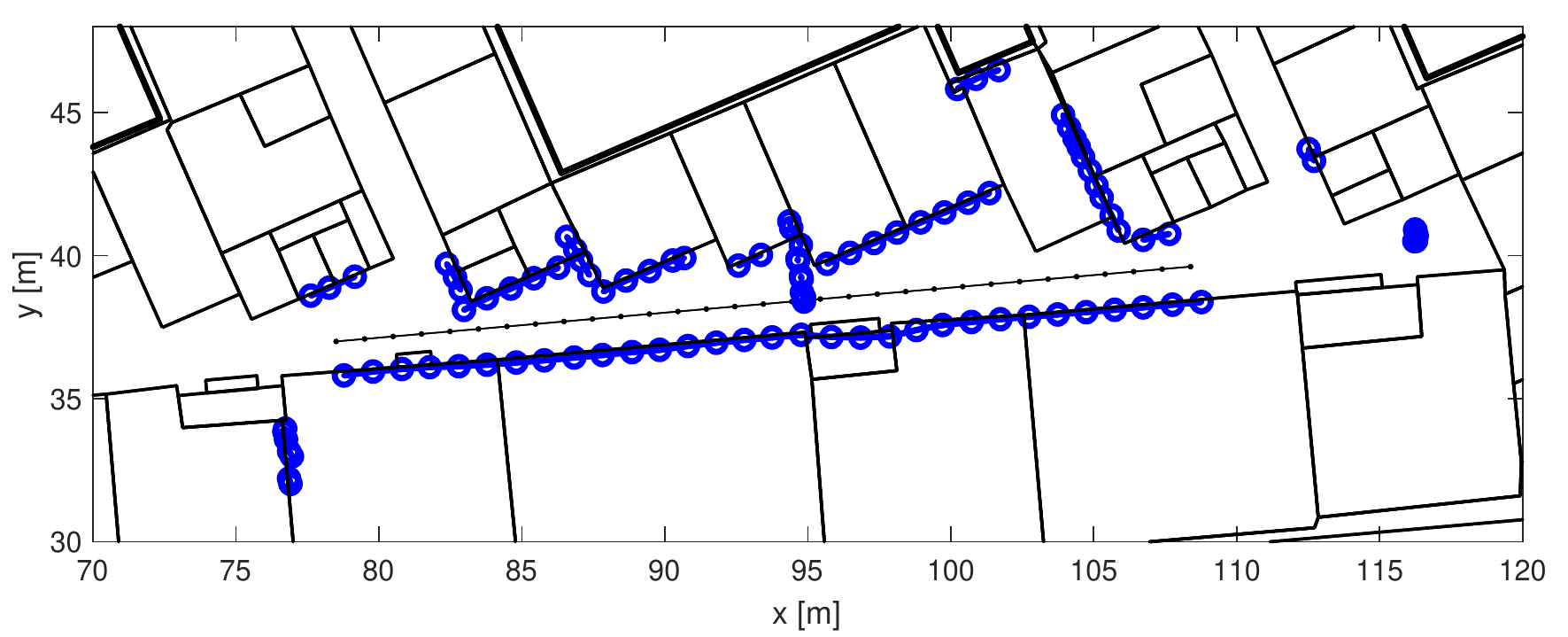}
        \label{fig:mapping_RT_IMM_forward}}
        \hfil
        \subfloat[Smoothed tracking-based dynamic mapping ]{\includegraphics[width=1\columnwidth]{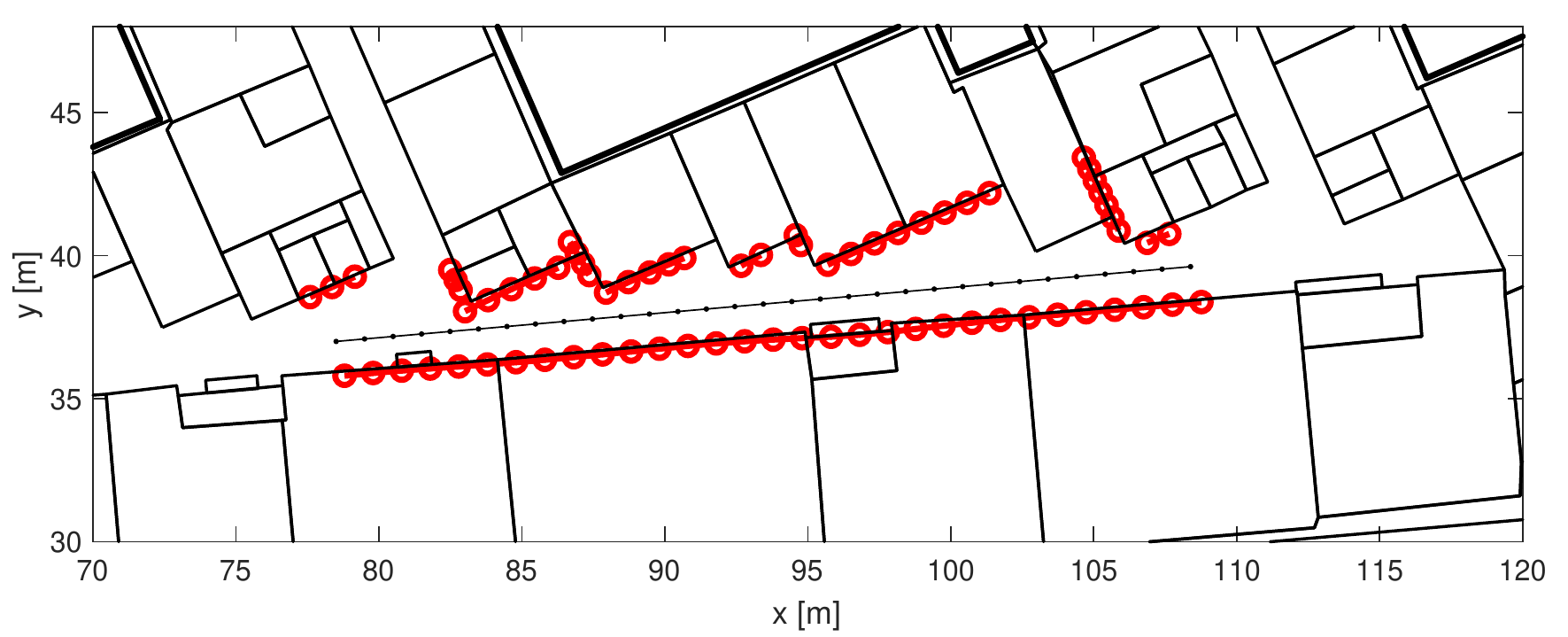}
        \label{fig:mapping_RT_IMM_smoothing}}
    \caption{Indoor mapping results with \emph{ray-tracing data} at 28~GHz, for {(a) ISTA-based static mapping}, (b) grid-based static mapping from \cite{myPaper_4_ICL20}, (c) tracking-based dynamic mapping without smoothing and (d) tracking-based dynamic mapping with smoothing.}
    \vspace{-3mm}
    \label{fig:mapping_RT}
\end{figure}

For validation purposes, the RF measurement campaign results are corroborated by realistic ray-tracing simulations using Wireless Insite\textregistered~\cite{InSite} software.  In these simulations, reproducing the scenario shown in Fig.~\ref{fig:setup}\subref{fig:setup_scenario}, the measurement device follows a similar trajectory as in the RF measurement campaign with 31 test locations {$1$~m apart} as shown in Fig.~\ref{fig:mapping_RT}, {using an angular scanning range from $-180^\circ$ to $180^\circ$}. The TX and RX array operation is simulated through a directive beam pattern with 3~dB beam-width of $17^\circ$, similar to the real directive antenna systems used in the RF measurements. In addition, the same antenna separation and height are used, and the carrier frequency is $28$~GHz.

The ray-tracing engine is configured to consider a maximum of 15 rays per simulation, and the number of allowed interactions is limited to six reflections and one diffraction. The walls, floor and ceiling are built using the frequency-specific materials, namely, ITU layered dry wall and floor or ceiling board. {The Lambertian diffuse scattering model is applied with a scattering factor of $0.2$ and a cross-polarization fraction of $0.4$ to all building materials except for glass. This way, the diffuse scattering compensates for the reasonable simplifications in the environment modeling}, compared to the true physical environment shown in Fig.~\ref{fig:setup}\subref{fig:setup_scenario}, that were allowed to reduce the computational complexity and simulation time. However, like the results will also illustrate, the ray-tracing environment {models the physical environment accurately}.

\begin{figure}[!t]
    \centering
        \subfloat[Range--angle chart via LS approach]{\includegraphics[width=0.90\columnwidth]{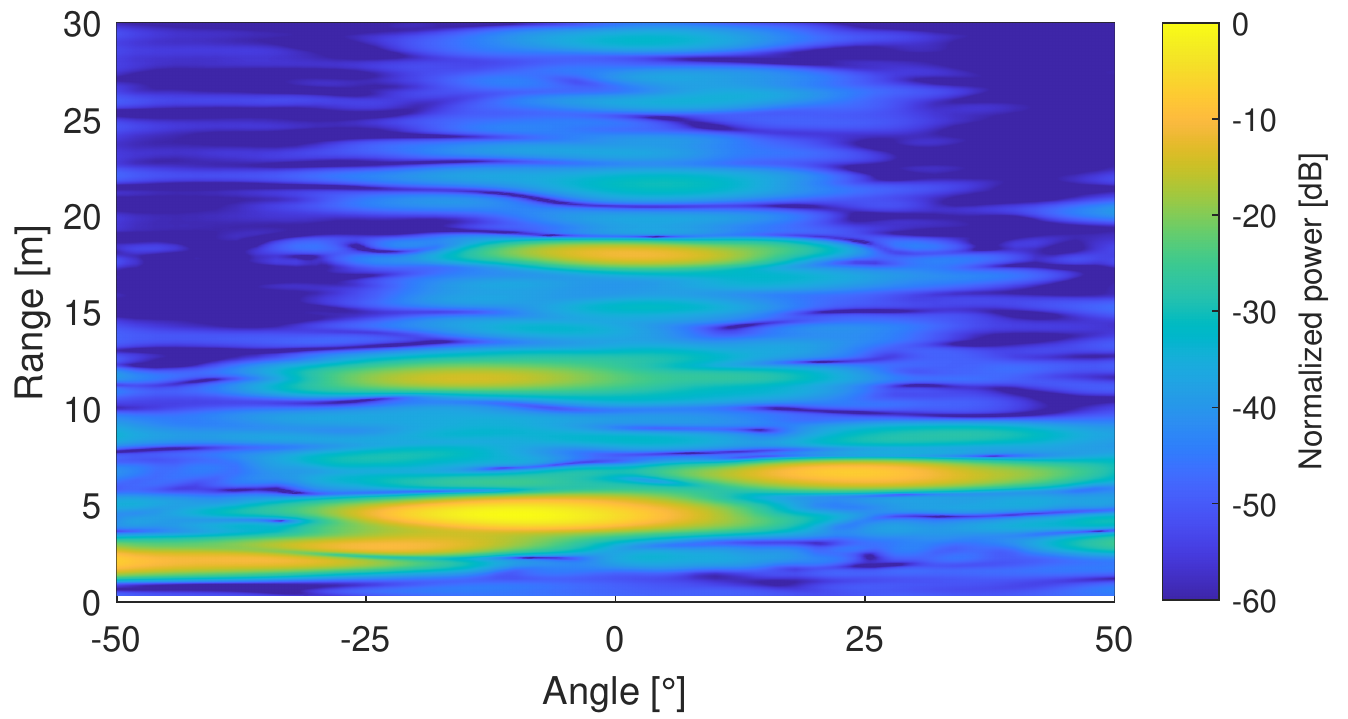}
        \label{fig:rangleAngleMap_LS}}
        \hfil
        \subfloat[{Range--angle chart via ISTA approach}]{\includegraphics[width=0.90\columnwidth]{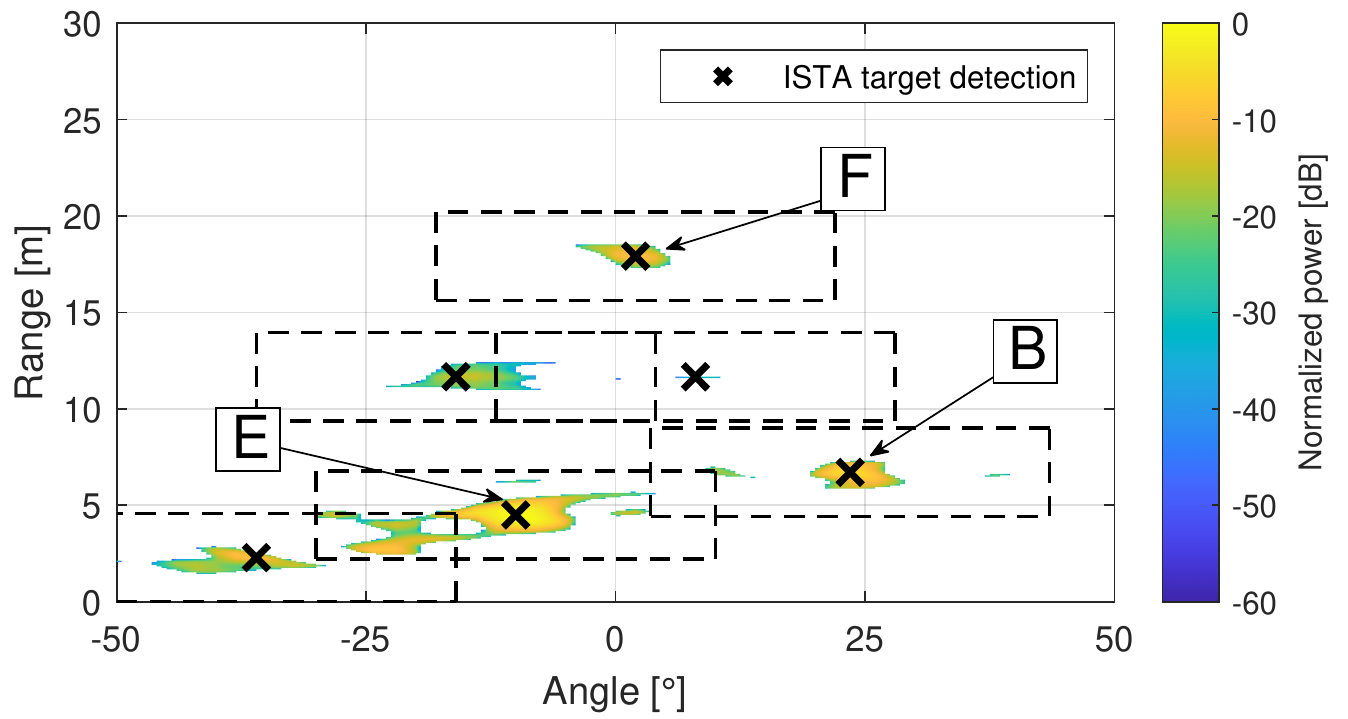}
        \label{fig:rangleAngleMap_CS}}
    \caption{Example range--angle charts obtained via (a) the LS approach and (b) the ISTA approach, when processing the 28~GHz  RF measurement data at location highlighted in green in Fig.~\ref{fig:result_maps}\protect\subref{fig:L2R_smoothing}.  {The targets \textit{B}, \textit{E} and \textit{F} defined along Fig.~\ref{fig:setup}\protect\subref{fig:setup_scenario} are highlighted in the range--angle chart for reference. {The dashed boxes illustrate the considered minimum target separation in range-angle domain utilized in the actual target detection.}}}
    \label{fig:rangleAngleMap}
\end{figure}

\begin{figure}[!t]
    \centering
        \includegraphics[width=1\columnwidth]{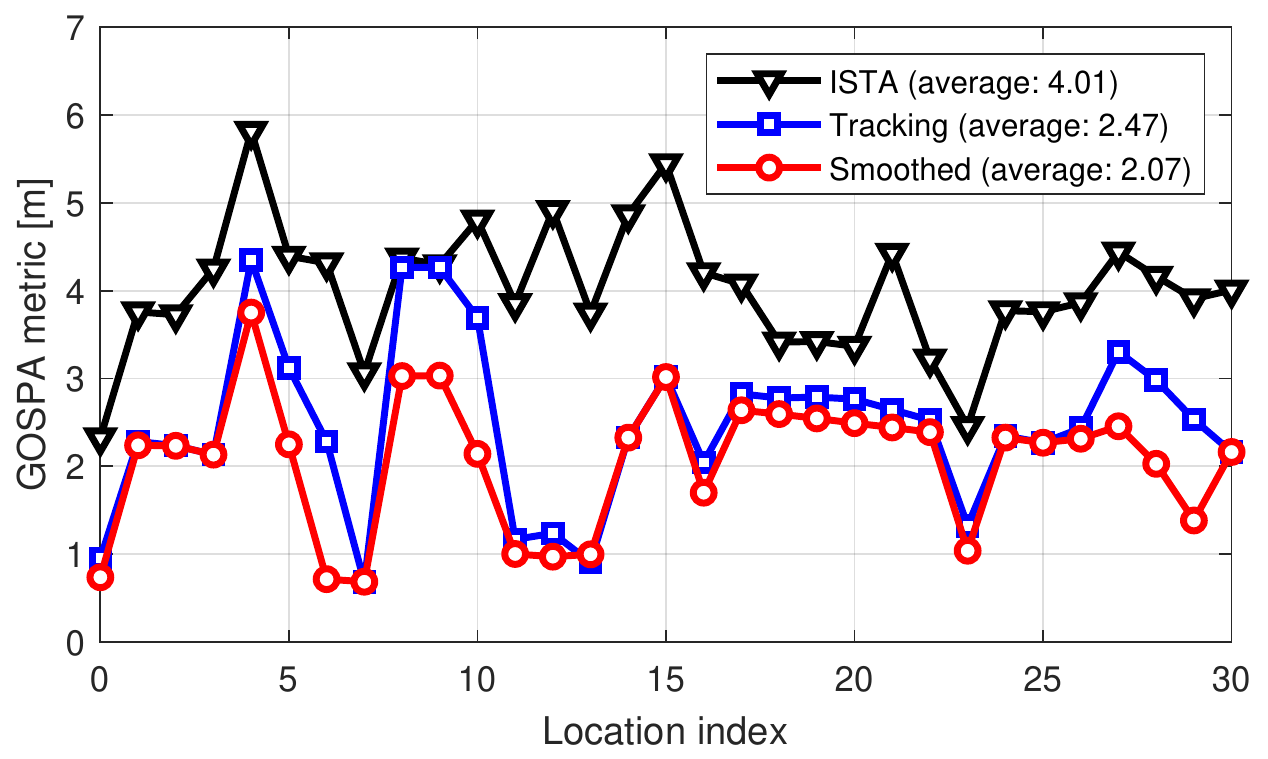}
     \vspace{-5mm}
    \caption{{GOSPA based quantitative performance assessment and comparison of the proposed methods for \textit{ray-tracing} data at 28~GHz.}}
    \label{fig:GOSPA_RT}
\end{figure}

\begin{figure*}[t]
    \centering
        \subfloat[Grid-based static mapping \cite{myPaper_4_ICL20} (moving from left to right) ]{\includegraphics[width=1\columnwidth]{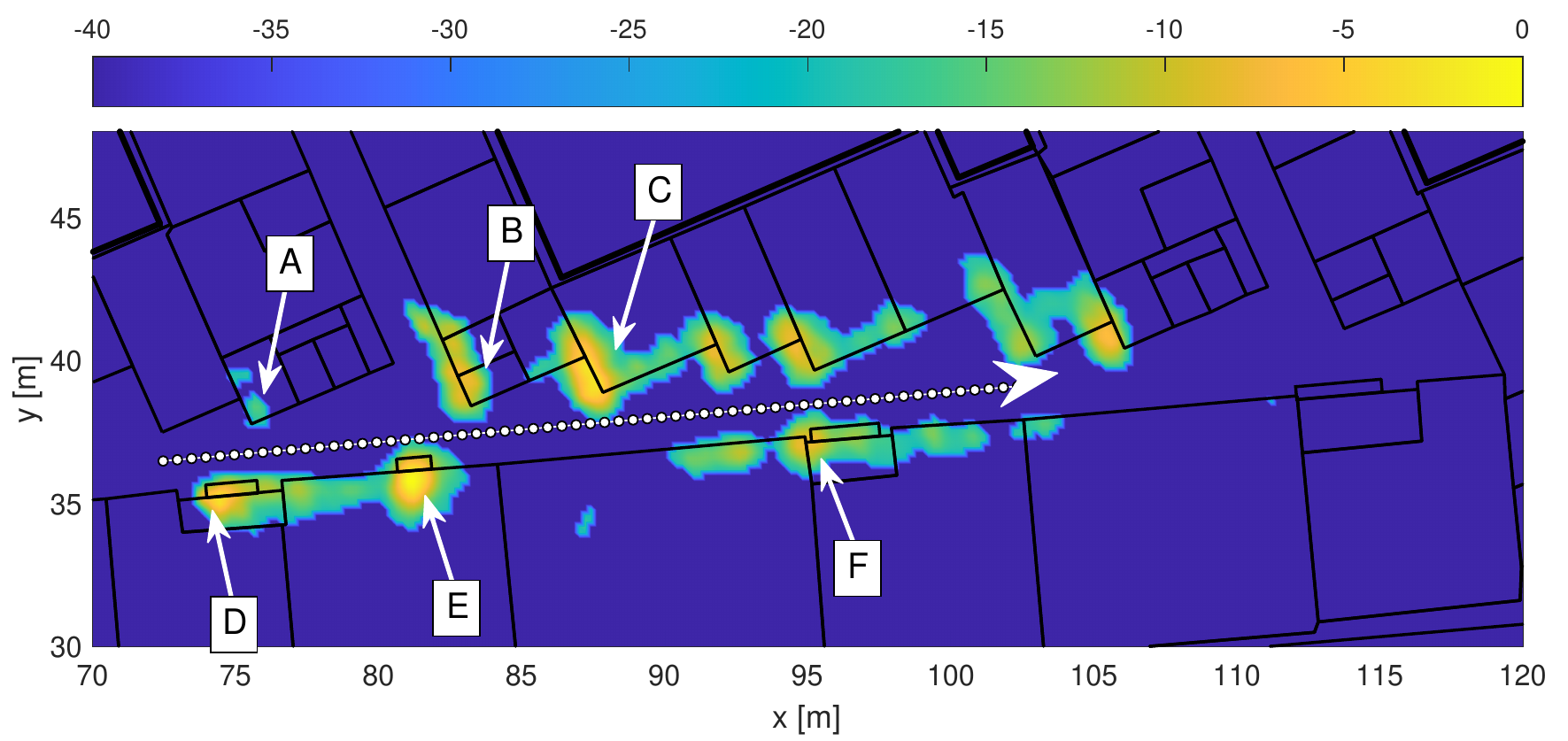}
        \label{fig:L2R_grid}}
        \subfloat[Grid-based static mapping \cite{myPaper_4_ICL20} (moving from right to left) ]{\includegraphics[width=1\columnwidth]{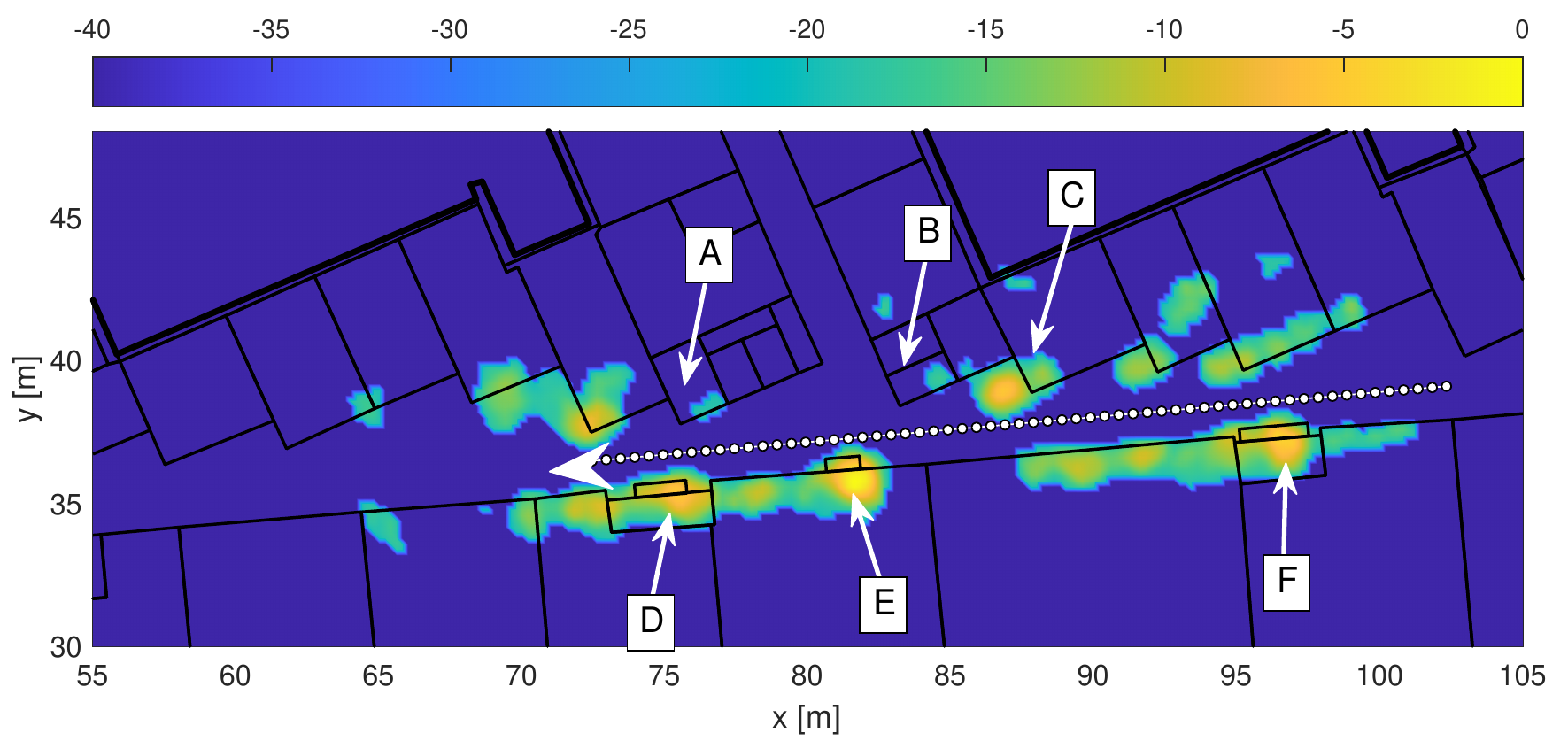}
        \label{fig:R2L_grid}}
        \hfil
        \subfloat[Tracking-based dynamic mapping (moving from left to right)]{\includegraphics[width=1\columnwidth]{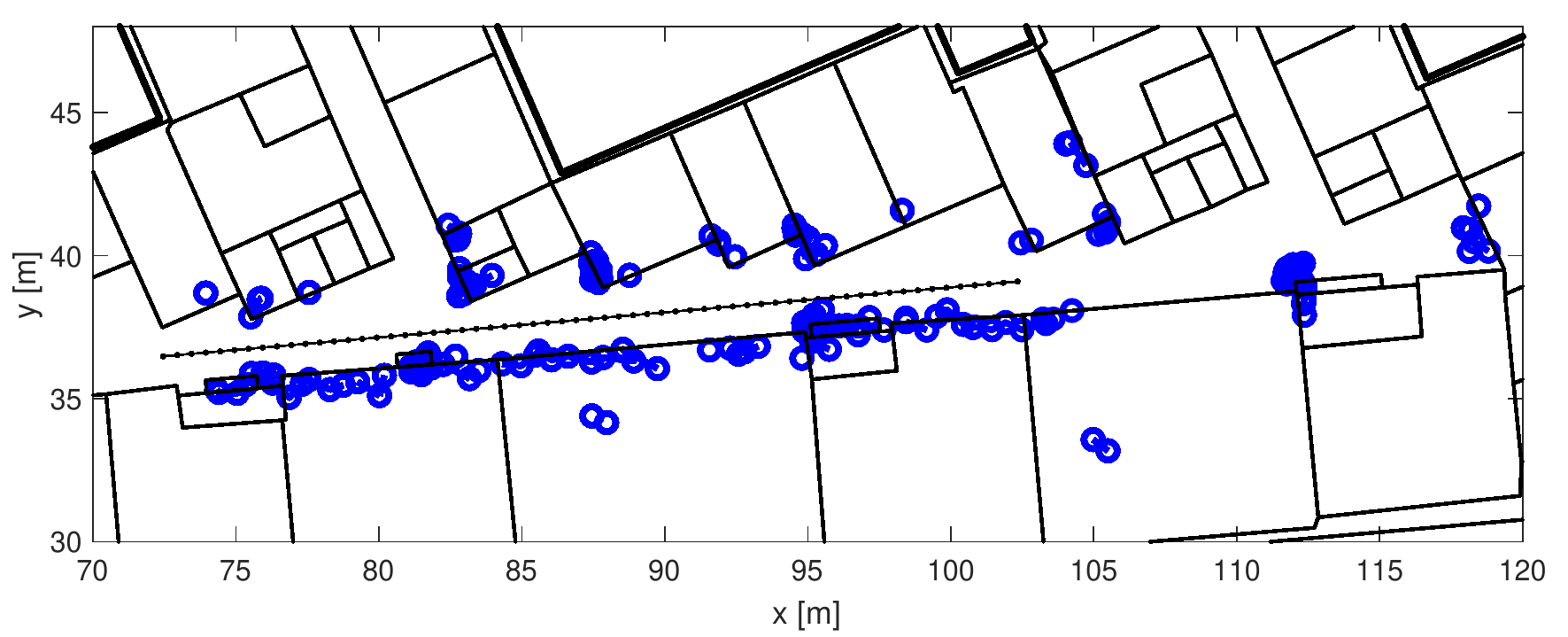}
        \label{fig:L2R_filter}}
        \subfloat[Tracking-based dynamic mapping (moving from right to left)]{\includegraphics[width=1\columnwidth]{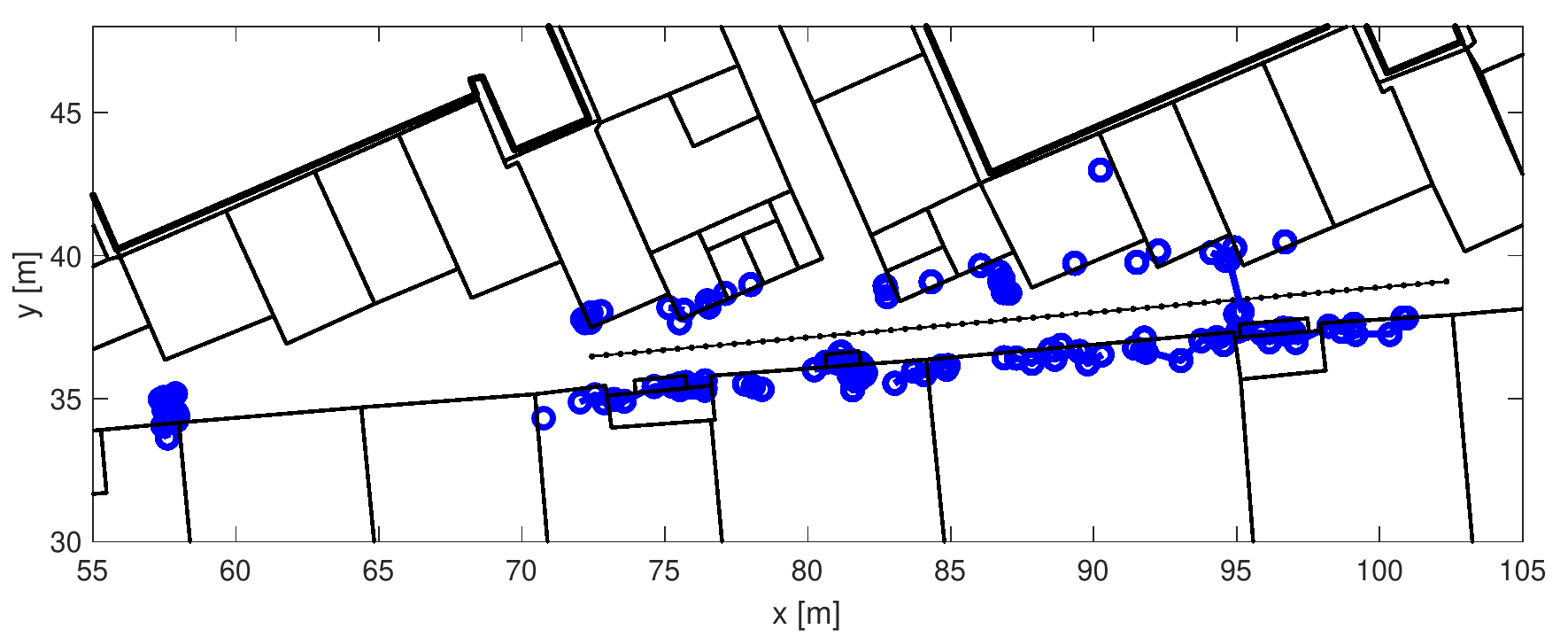}
        \label{fig:R2L_filter}}
        \hfil
         \subfloat[Smoothed tracking-based dynamic mapping (moving from left to right) ]{
         \includegraphics[width=1\columnwidth]{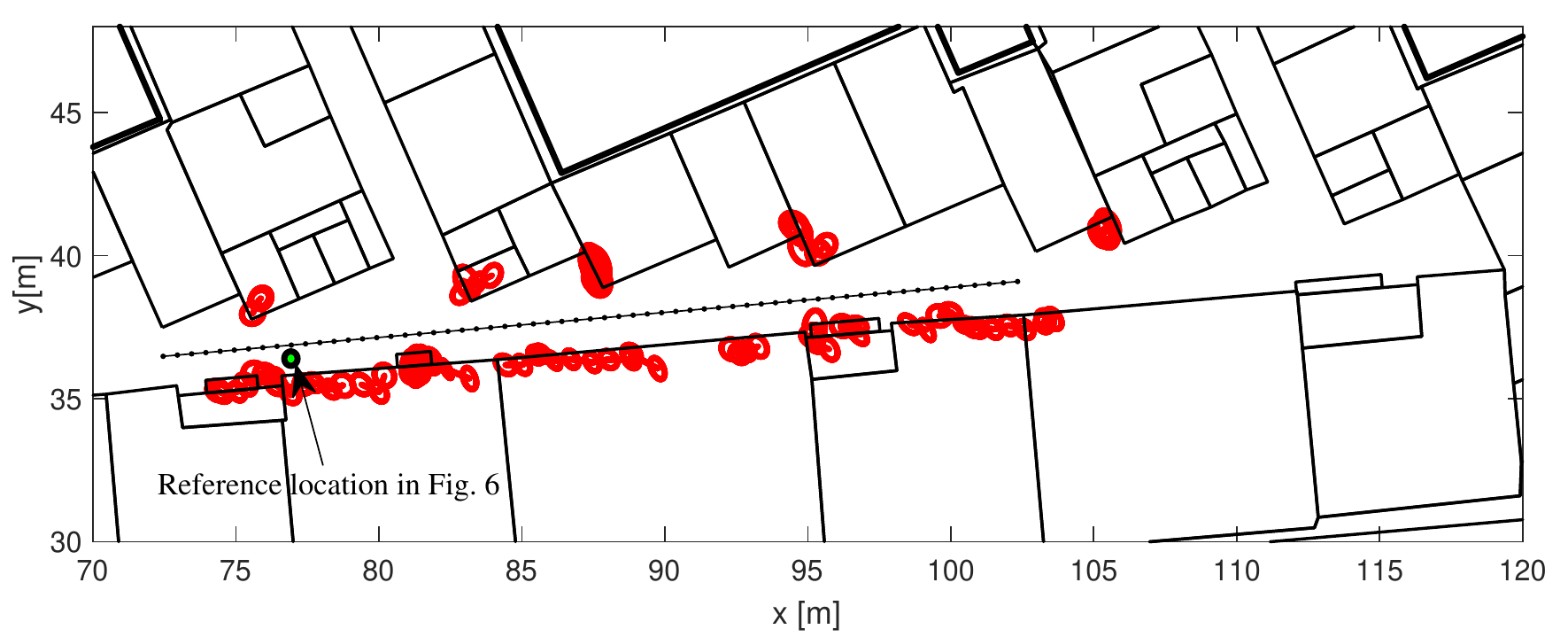}
        \label{fig:L2R_smoothing}}
        \subfloat[Smoothed tracking-based dynamic mapping (moving from right to left)]{        \includegraphics[width=1\columnwidth]{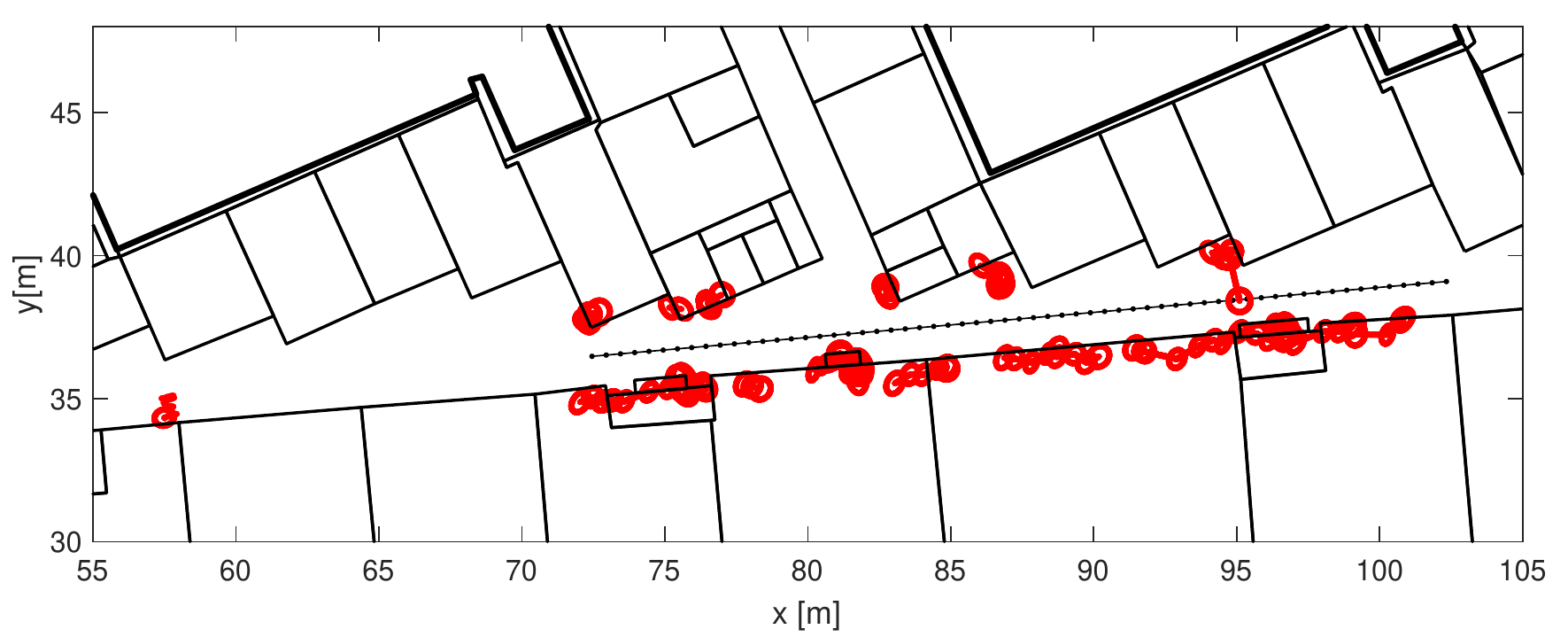}
        \label{fig:R2L_smoothing}}
    \caption{Indoor mapping results with \emph{RF measurement data} at 28~GHz. Subfigures (a), (c) and (e) illustrate different mapping methods for the center trajectory when the measurement equipment is moving from left to right along the corridor. Similarly, subfigures (b), (d) and (f) present mapping results when the measurement system senses the corridor while moving from right to left. }
    \vspace{-2mm}
    \label{fig:result_maps}
\end{figure*}

\vspace{-2mm}
\subsection{RF Measurement Equipment}

In the actual RF measurements, we use state-of-the-art mm-wave equipment to emulate the UE operation at the 28~GHz band, with selected equipment shown in Fig.~\ref{fig:setup}\subref{fig:setup_equipment}.
The baseline hardware platform in the measurement setup is the NI vector signal transceiver (PXIe-5840) which implements the RF TX and RX functionalities at intermediate frequency of 3.5~GHz, as well as controls the rest of the devices. In addition, two signal generators (Keysight N5183B--MXG) are used as local oscillators, which together with external mixers ({Marki Microwave T3-1040}) up- and down-convert the IF signal to and from the desired mm-wave carrier frequency.

To emulate the UE's phased array beam-steering operation, two directive horn antennas (PE9851A-20) are utilized for the TX and RX sides. These antennas are mounted on mechanical steering systems which enable to steer and direct the horns in the whole azimuth plane very accurately.
According to the specifications, both horns provide a nominal gain of 20~dBi with a 3~dB beam width of 17$^{\circ}$. The antennas are placed at one meter above the floor level with a separation of 60~cm in order to avoid larger mutual coupling between TX and RX chains. In the TX side, two external power amplifiers (PAs) are also used that together with the antenna system facilitate an EIRP of around +20~dBm. 
{In the RF measurements, 61 observation locations are considered as shown in Fig.~\ref{fig:result_maps} with an angular scanning range from $-50^\circ$ to $50^\circ$ due to hardware limitations.}

\vspace{-3mm}
\section{Results}

In this section, we provide the ray-tracing and RF measurement based results. We start with a short example of range--angle processing, utilizing the RF measurement data, while then put most of the focus on the actual mapping results, covering both the ray-tracing data and the RF measurement data. {Four different alternatives are considered and shown for the actual mapping results, namely, ISTA-based static mapping, grid-based static mapping from \cite{myPaper_4_ICL20}, tracking-based dynamic mapping and smoothed tracking-based dynamic mapping. 
In the ISTA-based static mapping, the sparse range--angle charts (see Sub-section~\ref{sec:RangeAngleProc}) from different sensing locations, together with 
the transformation in \eqref{eq:X_Y_coordinates}, are collected to obtain an elementary estimate of the map.
Similarly, the grid-based static mapping method from \cite{myPaper_4_ICL20} uses the ISTA-based charts as a reference while deploys then additional smoothing and thresholding on top of them.}

\vspace{-3mm}
\subsection{Example Range--Angle Processing Results}
We start by demonstrating the proposed range--angle processing technique described in Section~\ref{sec:rangeAngle_processing} using the 28~GHz RF measurement data. Specifically, the 10$^\text{th}$ location shown in Fig.~\ref{fig:result_maps}\subref{fig:L2R_smoothing} is considered as a representative example.
First, Fig.~\ref{fig:rangleAngleMap}\subref{fig:rangleAngleMap_LS} illustrates the resulting range--angle chart when applying the considered LS estimation approach. In this case, distances up to 30~$\text{m}$ and azimuth angles between {$-50^\circ$ to $50^\circ$} are investigated, using a total of $C_{\text{R}}=391$ and $\Nphi=221$ range and angle cells, respectively. In addition, a Hamming window is used to improve the side-lobe suppression in the range domain.
As it can be observed, the 400-MHz bandwidth used in the measurements facilitates high-accuracy range resolution, that enables to distinguish targets with mutual distances down to around 0.4~m. However, we also identify significant side-lobes in the angular-domain due to the TX and RX horn antenna patterns that degrade the target separation in this domain. 

Then, a sparse representation of this range--angle chart is obtained by applying the proposed ISTA approach {with regularization parameter of $\lambda=0.08$}, summarized in Algorithm~\ref{alg_ista}, to better facilitate the subsequent mapping phases. The corresponding results are shown in Fig.~\ref{fig:rangleAngleMap}\subref{fig:rangleAngleMap_CS}, {including also the actual detected targets that are used as input in the tracking-based dynamic mapping. In this case, a minimum target separation of $2.3~\text{m}$ and $20^\circ$ was considered for the range and angle domains, respectively, with a maximum of $10$ detected targets per location and considering an additional dynamic range related threshold of $-60~\text{dB}$.} As it can be seen, only the most significant target reflections remain in the sparse chart, suppressing the possible side-lobes in both range and angular domains. Hence, the ISTA approach is deployed as the main range--angle chart processing engine in the forthcoming mapping results.

\vspace{-3mm}
\subsection{Ray-Tracing based Mapping Results}\label{sec:RayTracing}

The considered mapping methods
are next assessed and validated with the ray-tracing based evaluations. As already noted, the ray-tracing environment resembles the true physical environment and the corresponding RF measurements as closely as possible -- with an exception that we adopt a wider angular scanning range from $-180^\circ$ to $180^\circ$ in order to seek for the maximum mapping performance. The obtained results with the different mapping methods are presented and illustrated in Fig.~\ref{fig:mapping_RT}, including also the building floor plan for reference.

{First, the ISTA-based static mapping approach is shown in Fig.~\ref{fig:mapping_RT}\subref{fig:mapping_RT_ISTA} and compared with the grid-based static mapping of Fig.~\ref{fig:mapping_RT}\subref{fig:mapping_RT_grid}.
As it can be observed, the ISTA-based approach is able to recover the most significant targets, providing a similar performance to the grid-based static method. 
Specifically, Fig.~\ref{fig:mapping_RT}\subref{fig:mapping_RT_grid} illustrates the final grid-based map after applying the different averaging, filtering and thresholding stages described in \cite{myPaper_4_ICL20} with the ISTA-based range--angle charts as the input.}
In the grid-based method, we pursue a 2D map of the simulated corridor using square cells with size of $0.2 \times 0.2 ~\text{m}^2$, which is consistent with the range resolution of about 0.4~m, to create the initial average map.
Next, as described in \cite{myPaper_4_ICL20}, a Gaussian kernel matrix with parameters $U=V=5$ and $\sigma=1$ is applied (see Eqs. (7)-(9) in \cite{myPaper_4_ICL20}) to smooth the map and reduce the effects of noise.
Finally, a thresholding stage 
is deployed to emphasize the most relevant targets of the environment.
As it can be observed, {both methods are able} to fairly accurately sense the indoor scenario, providing a mapping reconstruction that clearly reflects the true layout. 

Next, the same ray-tracing scenario and data is processed with the proposed tracking-based dynamic mapping approach, described in Section~\ref{sec:trackingBasedDynamicMapping}, {while considering the same ISTA target detection parameters as in the previous subsection}. The corresponding results are shown in Fig.~\ref{fig:mapping_RT}\subref{fig:mapping_RT_IMM_forward} and \subref{fig:mapping_RT_IMM_smoothing}, without and with IMM smoothing, respectively. These results consider the IMM-based method which deploys two separate EKF filters using the CWNV and CWNA motion models for stationary and moving scatterers, respectively.
For the measurement selection procedure, we have chosen the single interaction model probability of $p^{\text{prior}}({\zeta}_{l,k} =\mathcal{H}_0) = 0.5$ and threshold values of $p^\textrm{sample}_\textrm{th} = 0.01$ and $d_\textrm{th} = 2.3~m$. In addition, the standard deviations for the single-reflection model and the multi-reflection model are defined as $\sigma_{\mathcal{H}_0}=2$ and $\sigma_{\mathcal{H}_1}=10$.
For the initial state covariance in the EKF models, we use a variance of $10^4~m^2$ for the target x and y coordinates, as well as  $0.5~{m^2}/{s^2}$ for the target velocities in x and y directions when considering the CWNA model. Moreover, we choose to emphasize tracking of single-order specular reflections, and thus we initialize the target velocities at the second time step by projecting the known UE movement with a measured target angle to the movement of an assumed single-order specular reflection point at a wall. The used power spectral densities of the state covariance matrices $Q_c$ of the CWNV and CWNA models are $10^{-6}~{m^2}/{s}$ and $0.05~{m^2}/{s^3}$, respectively. {The range and angle related measurement standard deviations are set to $10^{\circ}$ and $0.2~m$, respectively}. The initial probabilities of the scatterer motion models ($\mu^j_0$) for the CWNA and CWNV are set as {$0.15$ and $0.85$}, respectively.

{In the forward IMM filter pass, the scatterer position estimates are tracked together with their covariance matrices. At each step, they are updated using the newly acquired angle and distance measurements.} To this end, Fig.~\ref{fig:mapping_RT}\subref{fig:mapping_RT_IMM_forward} shows all the tracked scatterer positions through the UE trajectory. The positions corresponding to the same scatterer cross-identified between multiple time instants are interconnected forming trajectories that are shown to follow the walls of the corridor with notable accuracy. Note that the covariance matrices of the scatterer positions become smaller as more measurement are available, hence, the position estimates in the beginning of tracking are not as reliable as those closer to the end. Also, we can still observe a few apparent scatterers beyond the walls, corresponding, e.g., to double reflections that have passed the measurement selection stage. 

After the UE has finished the tracking process, it provides the tracked targets for post-processing, i.e., to the IMM smoother that is deployed to further improve the final map representation.
During the backwards IMM smoothing pass, the scatterers' position and covariance estimates are updated taking into account all measurements available at the last tracking time instant. In the final map processing step, we use the smoothed covariance of each position estimate as a measure of reliability, allowing thus to extract only the most reliable ones. The corresponding final map
shown in Fig.~\ref{fig:mapping_RT}\subref{fig:mapping_RT_IMM_smoothing}, 
demonstrates that a number of scatterers did not pass to the final stage due to their large covariance, especially those in the beginning of the tracking process, those far away from the TX and those corresponding to the multiple reflections.

{Utilization of ray-tracing data enables extraction of ground-truth information, including the interaction coordinates of single-bounce radio paths at different walls. This allows quantitative performance evaluation of the studied methods by comparing the estimated scatterer positions with the available ground-truth data. In Fig. \ref{fig:GOSPA_RT}, the performance of ISTA-based measurements (including the ISTA target detection), forward IMM tracking, and backwards IMM smoothing are shown in terms of GOSPA metric \cite{rahmathullah_2017}, which considers the achieved estimation accuracy as well as the number of missed detections and false detections. The proposed IMM smoother delivers the best performance by providing almost 50\% lower GOSPA compared to the ISTA-based measurements. The observed fluctuation in GOSPA over different time steps originates from the variation of measurable walls, as the UE moves along the corridor. Whenever measurements from a new wall become available, the IMM filters need to be initialized. Especially with short walls, this is a challenge as the filter states might not have enough time to properly converge. Nonetheless, the results show that despite these potential challenges, the proposed IMM approach is able to considerably improve the mapping performance compared to the original ISTA measurements.}

\vspace{-2mm}
\subsection{RF Measurement based Mapping Results}

Finally, the proposed methods are assessed and tested with RF measurements to validate the considered sensing and mapping functionality with real-world equipment and physical environment. We address how the mapping system performance is subject to the UE orientation by showing mapping results for two main UE trajectories separately. To this end, Fig.~\ref{fig:result_maps}\subref{fig:L2R_grid}, \subref{fig:L2R_filter} and \subref{fig:L2R_smoothing} compare the grid-based static and the tracking-based dynamic mapping results when the measurement equipment is moving along the corridor from left to right, while Fig.~\ref{fig:result_maps}\subref{fig:R2L_grid}, \subref{fig:R2L_filter} and \subref{fig:R2L_smoothing} show the corresponding mapping results for the opposite moving direction, from right to left. 

It can be seen how the proposed dynamic mapping method accurately reconstructs the complex physical environment despite the fairly limited scanning range of {$-50^\circ$ to $50^\circ$} of the available antenna systems, compared to the ray-tracing scenario, and despite the numerous real-world effects and impairments that are present in the measurement data.
We especially highlight the locations of the main corridor walls (marked with \textit{A}, \textit{B} and \textit{C}) and the metallic lockers (marked with \textit{D}, \textit{E} and \textit{F}) locations, both in Fig.~\ref{fig:setup} and in Fig. \ref{fig:result_maps}, for better interpretation of the results.
We can observe how the tracking-based dynamic approach is able to accurately track the main targets of the environment, and subsequently, improve the map quality -- especially when the IMM smoothing stage is deployed. 

\begin{figure}[!t]
    \centering
      \includegraphics[width=1\columnwidth]{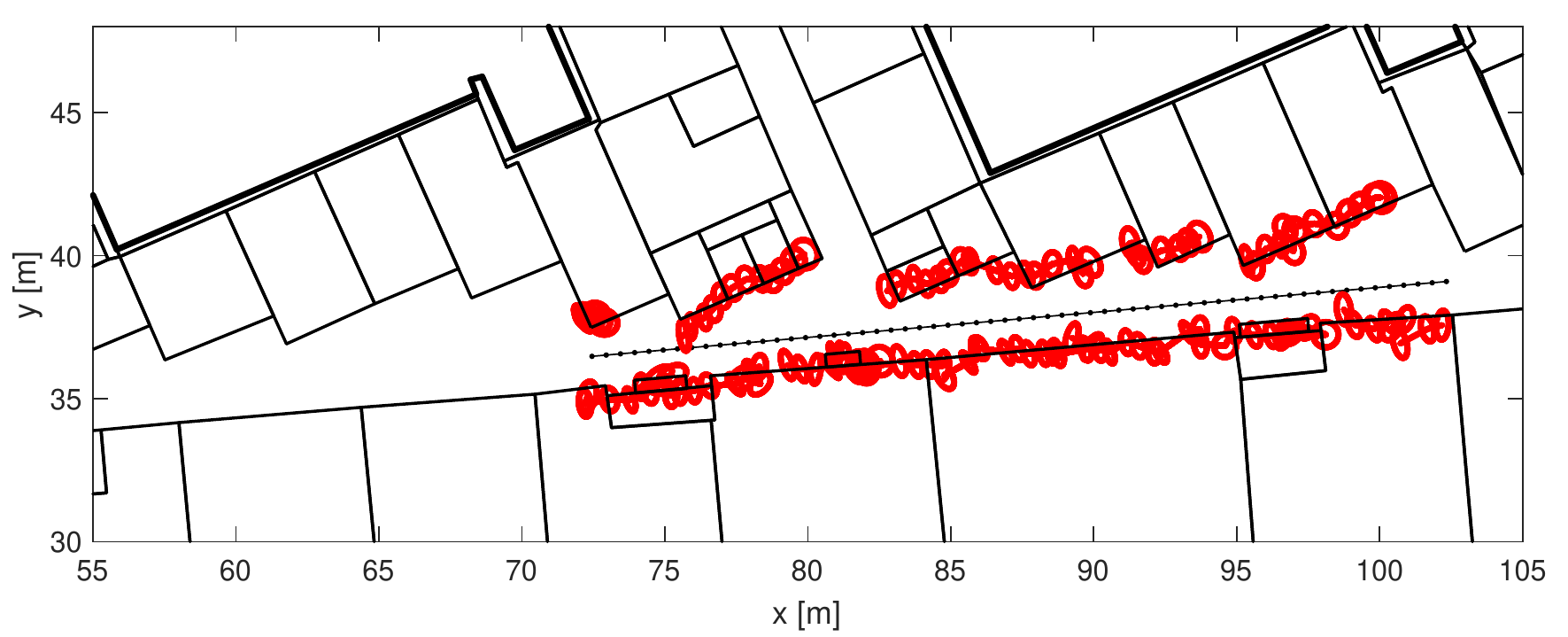}
    \vspace{-5mm}
    \caption{{Smoothed tracking-based dynamic mapping with \emph{RF measurement data} for wider scanning range of $-90^\circ$ to $90^\circ$. The measurement equipment is moving from right to left along the corridor.}}
    \label{fig:R2L_smoothing_widerScanning}
\end{figure}

Finally, we showcase that by increasing the scanning range of the measurement setup, a more rich and further accurate representation of the environment can be obtained. In this regard, Figure~\ref{fig:R2L_smoothing_widerScanning} illustrates the smoothed tracking-based dynamic mapping results when a wider scanning range of $-90^\circ$ to $90^\circ$ is used in the measurements.
As it can be observed, the wider scanning range provides a more complete representation of the indoor map in comparison with the map shown in Fig.~\ref{fig:result_maps}\subref{fig:R2L_smoothing}. 
Overall, the mapping results with real-world RF measurement data demonstrate the applicability of the proposed methods also in true complex physical environments, allowing to extract situational awareness in an efficient manner.
{In general, when the future mobile networks further evolve towards the 100~GHz bands, and perhaps even beyond, the relative role of diffuse scattering is likely to increase \cite{journal_40}. The novel state formulation described in this article -- taking into account both the specular and diffuse components -- combined with the IMM filtering framework is offering a versatile tool for high-accuracy radio environment mapping also in such networks. Demonstrating that through concrete RF measurements is one important ingredient in our future work.}

\section{Conclusions}
\label{sec:Conclusion}
This paper investigated the sensing and environment mapping prospects of mm-wave 5G NR and beyond networks with specific emphasis on the UE side for mobile mapping applications. In the considered framework, the UE operates as a joint communication and sensing system, scanning its surrounding environment while steering its beam pattern towards different directions and observing the target reflections. First, we derived a novel LS-based processing technique to estimate sparse range--angle charts to facilitate the subsequent mapping processing. Then, an advanced dynamic mapping approach was presented, building on novel state model with diffuse and specular scattering together with IMM-EKF filtering and smoothing. 
While the sparse range--angle charts already allow for baseline static mapping capability, the dynamic tracking-based approach that builds on the IMM-EKF processing solution incorporating both specular reflections and diffuse scattering allows to track individual scatterers over time. This together with the IMM smoothing and the described novel scatterer measurement selection and association methods provide an efficient framework to reconstruct and map complex physical environments.
The applicability of the proposed methods and algorithms was then assessed and evaluated, through both ray-tracing simulations and actual RF measurements at the 28~GHz band in practical indoor office type of an environment. The obtained results demonstrate the good applicability of the proposed methods, while the IMM-EKF based dynamic tracking solution was shown to clearly outperform the more simple static approaches. Our future work will consider extensions to 3D sensing and mapping, while extending also to the actual SLAM where the coordinates of the sensing device are also unknown. {Additionally, the future work contains generalizing the presented methods for multiple simultaneous transmit and/or receive beams, and carrying out measurements at 60--100~GHz bands.}  

\begin{appendices}
 
 \section{LS Solution}\label{app_ls}
The LS approach to estimate the range--angle chart $\bbig$ in \eqref{eq_obs} is formulated as
\begin{align}
 \min_{\bbig} \sum_{m=0}^{M-1} \normsmall{\ybig_m - \xbig_m \odot \cbig \bbig \gbig  }_F^2 ~.
\end{align}
After vectorization, we have
\begin{align}\label{eq_bb_ls}
 \min_{\bb} \sum_{m=0}^{M-1} \normsmall{\yy_m - \xx_m \odot (\gbig^T \otimes \cbig) \bb }_2^2 ~,
\end{align}
where $\xx_m \triangleq \vecc{\xbig_m}$, $\yy_m \triangleq \vecc{\ybig_m}$, $\bb \triangleq \vecc{\bbig}$ {and $\vecc{\cdot}$ denotes the vectorization}. Further simplifying \eqref{eq_bb_ls} yields
\begin{align}\label{eq_bb_ls2}
 \min_{\bb} \sum_{m=0}^{M-1} \normsmall{\yy_m - \boldPhi_m \bb }_2^2 ~,
\end{align}
where
\begin{align}\label{eq_phim}
    \boldPhi_m \triangleq \diag{\xx_m} \left(\gbig^T \otimes \cbig\right) \in \complexset{N I}{C_{\text{R}} \Nphi},
\end{align}
is a known matrix. Defining then
\begin{align} \label{eq_y_all}
    \yy &= \left[ \yy_0^T, \, \ldots , \, \yy_{M-1}^T \right]^T \in \complexset{NIM}{1} ~, \\ \label{eq_boldphi}
    \boldPhi &= \left[ \boldPhi_0^T, \, \ldots , \, \boldPhi_{M-1}^T \right]^T \in \complexset{NIM}{C_{\text{R}} \Nphi} ~,
\end{align}
\eqref{eq_bb_ls2} becomes
\begin{align}\label{eq_bb_ls3}
 \min_{\bb} \normsmall{\yy - \boldPhi \bb }_2^2 ~.
\end{align}
The LS estimate of the range--angle chart in \eqref{eq_bb_ls3} can thus be obtained as
\begin{align}\label{eq_ls_sol}
    \bbhatls = \boldPhi^\dagger \yy ~,
\end{align}
where $(\cdot)^\dagger$ denotes matrix pseudo-inverse. {In the regime of large number of OFDM subcarriers/symbols,} the LS solution can be approximated as (see Appendix~\ref{app_psk} for the derivations):
\begin{align}\label{eq_bhat_ls2}
    \bbighatls {\approx} \underbrace{ \left( \cbig^H \cbig\right)^{-1} \cbig^H }_{\text{Range~compression}} \underbrace{ \frac{1}{M}\sum_{m=0}^{M-1}( \xbig_m^{*} \odot \ybig_m ) }_{\text{Coherent~integration}} \underbrace{ \gbig^H \left( \gbig \gbig^H \right)^{-1} }_{\text{Pattern~correlation}} ~,
\end{align}
which consists of coherent integration over the $M$ received symbols, range compression/matched filtering and antenna pattern correlation operations.

     \section{Approximated LS Solution}\label{app_psk}

In this {Appendix}, we {show how \eqref{eq_ls_sol} can be approximated as \eqref{eq_bhat_ls2}} 
using the Kronecker structure in \eqref{eq_phim}. Suppose $\boldPhi$ has full column rank -- an assumption that is well-justified due to randomness of data symbols $\xx_m$ and large number of OFDM subcarriers/symbols so that $N I M > C_{\text{R}} \Nphi$. Then, we can rewrite \eqref{eq_ls_sol} as
\begin{align} \nonumber
    \bbhatls &= (\boldPhi^H \boldPhi)^{-1} \boldPhi^H \yy \\ \nonumber
    &= \left( \sum_{m=0}^{M-1} \boldPhi_m^H \boldPhi_m \right)^{-1} \sum_{m=0}^{M-1} \boldPhi_m^H \yy_m \\ \nonumber
    &= \left[ \left(\gbig^{*} \otimes \cbig^H\right) \sum_{m=0}^{M-1} \diag{\xx_m}^H  \diag{\xx_m} \left(\gbig^T \otimes \cbig\right) \right]^{-1} \\ \nonumber & ~~~~\times \left(\gbig^{*} \otimes \cbig^H\right) \sum_{m=0}^{M-1} \diag{\xx_m}^H \yy_m \\ \nonumber
    &= \left[ \left(\gbig^{*} \otimes \cbig^H\right) \sum_{m=0}^{M-1} \diag{\{\lvert x_{i,m,n} \rvert^2\}_{i,n}} \left(\gbig^T \otimes \cbig\right) \right]^{-1} \\ \label{eq_ls_approx1} & ~~~~\times \left(\gbig^{*} \otimes \cbig^H\right) \sum_{m=0}^{M-1} (\xx_m^{*} \odot \yy_m) \\ \nonumber
    &\approx \left[ \left(\gbig^{*} \otimes \cbig^H\right)  M \Imatrix \left(\gbig^T \otimes \cbig\right) \right]^{-1} \\ \label{eq_ls_approx2} & ~~~~\times   \left(\gbig^{*} \otimes \cbig^H\right) \sum_{m=0}^{M-1} (\xx_m^{*} \odot \yy_m) ~.
\end{align}
Going from \eqref{eq_ls_approx1} to \eqref{eq_ls_approx2}, we use the law of large numbers for sufficiently large $M$ by assuming that the constellation has an average magnitude of $1$. The expression in \eqref{eq_ls_approx2} can be further re-written as
\begin{align}
    \bbhatls &{\approx} \left[ \left(\gbig^{*}  \gbig^T \right) \otimes \left( \cbig^H \cbig\right)  \right]^{-1} \left(\gbig^{*} \otimes \cbig^H\right) \\ \nonumber & ~~~~\times \frac{1}{M} \sum_{m=0}^{M-1} (\xx_m^{*} \odot \yy_m)
    \\ \nonumber
    &=  \left[ \left(\gbig^{*}  \gbig^T \right)^{-1} \otimes \left( \cbig^H \cbig\right)^{-1} \right] \left(\gbig^{*} \otimes \cbig^H\right) \\ \nonumber & ~~~~\times \frac{1}{M} \sum_{m=0}^{M-1} (\xx_m^{*} \odot \yy_m) \\ \nonumber
    &= \left[ \left(\gbig^{*}  \gbig^T \right)^{-1} \gbig^{*} \otimes \left( \cbig^H \cbig\right)^{-1} \cbig^H \right] \\ \nonumber & ~~~~\times \frac{1}{M} \sum_{m=0}^{M-1} (\xx_m^{*} \odot \yy_m) \\ \label{eq_ls_sol2}
    &= \veccs{ \left( \cbig^H \cbig\right)^{-1} \cbig^H \frac{1}{M}\sum_{m=0}^{M-1}( \xbig_m^{*} \odot \ybig_m ) \\ \nonumber & ~~~~\times \gbig^H \left( \gbig \gbig^H \right)^{-1}  } ~,
\end{align}
which finally yields
\begin{align}\label{eq_bhat_ls}
    \bbighatls {\approx} \left( \cbig^H \cbig\right)^{-1} \cbig^H \frac{1}{M}\sum_{m=0}^{M-1}( \xbig_m^{*} \odot \ybig_m ) \gbig^H \left( \gbig \gbig^H \right)^{-1} ~.
\end{align}

As seen from \eqref{eq_bhat_ls}, the LS solution corresponds to a matched filtering operation. In particular, the observation matrix is first multiplied by the conjugate of the OFDM transmit symbols to cancel out their effect. Then, the resulting matrices are coherently integrated over $M$ symbols relying on the assumption of negligible Doppler over the duration of $M$ symbols. Next, left multiplication by $\left( \cbig^H \cbig\right)^{-1} \cbig^H$ represents range compression/matched filtering via the range-dependent frequency-domain steering matrix $\cbig$. 
Notice that the columns of $\cbig$ coincide with those of an $N \times N$ DFT matrix. This means that matched filtering via $\left( \cbig^H \cbig\right)^{-1} \cbig^H$ is essentially a normalized IDFT operation over the columns of $\xbig_m^{*} \odot \ybig_m$, which is a standard approach in OFDM radars \cite{journal_13,braun2014ofdm}. Similarly, right multiplication by $\gbig^H \left( \gbig \gbig^H \right)^{-1}$ in \eqref{eq_bhat_ls} represents correlation/matched filtering with the antenna pattern. Hence, \eqref{eq_bhat_ls} can be interpreted as matched filtering in the range--angle domain, which essentially maximizes the SNR of the target detection.

\end{appendices}

\bibliographystyle{IEEEtran}
\bibliography{main}

\vspace{5mm}
\section*{Biographies}
\vspace{-16mm}
\begin{IEEEbiographynophoto}{Carlos Baquero Barneto}[S'18]
is a doctoral candidate at the Unit of Electrical Engineering at Tampere University, Finland. He received his B.Sc.\ and M.Sc.\ degrees in telecommunication engineering from Universidad Polit\'ecnica de Madrid, Spain, in 2017 and 2018, respectively. His research interest lies in the area of joint communication and sensing systems' design, with particular emphasis on 5G and beyond mobile radio networks.
\end{IEEEbiographynophoto}

\vspace{-16mm}
\begin{IEEEbiographynophoto}{Elizaveta Rastorgueva-Foi}
is a doctoral candidate at the Unit of Electrical Engineering at Tampere University, Finland, where she also received her M.Sc. degree in electrical engineering in 2019. Her research interests include statistical signal processing, positioning and location-aware communications in mobile networks with an emphasis on 5G and beyond systems.
\end{IEEEbiographynophoto}

\vspace{-16mm}
\begin{IEEEbiographynophoto}{Musa Furkan Keskin}
is a researcher and a Marie Skłodowska-Curie Fellow (MSCA-IF) in the department of Electrical Engineering at Chalmers University of Technology, Gothenburg, Sweden. He obtained the B.S., M.S., and Ph.D degrees from the Department of Electrical and Electronics Engineering, Bilkent University, Ankara, Turkey, in 2010, 2012, and 2018, respectively. His project "OTFS-RADCOM: A New Waveform for Joint Radar and Communications Beyond 5G" is granted by the European Commission through the H2020-MSCA-IF-2019 call. His current research interests include joint radar-communications, waveform design, and positioning in 5G and beyond 5G systems.
\end{IEEEbiographynophoto}

\vspace{-16mm}
\begin{IEEEbiographynophoto}{Taneli Riihonen}
[S'06, M'14] received his D.Sc.\ degree in electrical engineering (with honors) from Aalto University, Helsinki, Finland, in 2014. He is currently a tenure-track assistant professor at the Faculty of Information Technology and Communication Sciences, Tampere University, Finland. His research activity is focused on communication fundamentals as well as multicarrier, multiantenna, multihop and full-duplex wireless techniques with current interest in the evolution of beyond 5G radio systems.
\end{IEEEbiographynophoto}

\vspace{-16mm}
\begin{IEEEbiographynophoto}{Matias Turunen}
is a researcher and laboratory specialist at the Department of Electrical Engineering, Tampere University (TAU), Finland. His research interests include inband full-duplex radios with an emphasis on analog RF cancellation, OFDM radar, and 5G New Radio systems.
\end{IEEEbiographynophoto}

\vspace{-16mm}
\begin{IEEEbiographynophoto}{Jukka Talvitie}
[S'08, M'16] received the M.Sc. and D.Sc. degrees from Tampere University of Technology, Finland, in 2008 and 2016, respectively. He is currently a University Lecturer in the Faculty of Information Technology and Communication Sciences, Tampere University (TAU), Finland. His research interests include signal processing for wireless communications, radio-based positioning, radio link waveform design, and radio system design, particularly concerning 5G and beyond mobile technologies.
\end{IEEEbiographynophoto}

\vspace{-16mm}
%
\begin{IEEEbiographynophoto}{{Henk~Wymeersch}}
[S'01, M'05, SM'19] obtained the Ph.D. degree in Electrical Engineering/Applied Sciences in 2005 from Ghent University, Belgium. He is currently a Professor of Communication Systems with the Department of Electrical Engineering at Chalmers University of Technology, Sweden. He is also a Distinguished Research Associate with Eindhoven University of Technology. Prior to joining Chalmers, he was a postdoctoral researcher from 2005 until 2009 with the Laboratory for Information and Decision Systems at the Massachusetts Institute of Technology. Prof. Wymeersch served as Associate Editor for IEEE Communication Letters (2009-2013), IEEE Transactions on Wireless Communications (since 2013), and IEEE Transactions on Communications (2016-2018).  During 2019-2021, he was a IEEE Distinguished Lecturer with the Vehicular Technology Society.  His current research interests include the convergence of communication and sensing, in a 5G and Beyond 5G context. 
\end{IEEEbiographynophoto}

\vspace{-16mm}
\begin{IEEEbiographynophoto}{Mikko Valkama}
[S’00, M’01, SM’15, F’22] received his M.Sc. (Tech.) and D.Sc. (Tech.) degrees (both with honors) from Tampere University of Technology, Finland, in 2000 and 2001, respectively. In 2003, he was with the Communications Systems and Signal Processing Institute at SDSU, San Diego, CA, as a visiting research fellow. Currently, he is a full professor and the head of the Unit of Electrical Engineering at the newly formed Tampere University, Finland. His general research interests include radio communications, radio localization, and radio-based sensing, with particular emphasis on 5G and 6G mobile radio networks.
\end{IEEEbiographynophoto}

\end{document}